\pdfoutput=1
\documentclass{aa}  
\usepackage{graphicx}
\usepackage{amsmath} 
\usepackage{txfonts}
\usepackage[T1]{fontenc}
\usepackage{afterpage}
\usepackage{amssymb}
\usepackage{natbib} 
\bibpunct{(}{)}{;}{a}{}{,}
\newcommand{\Ms}{\ensuremath{M_\odot}}
\newcommand{\Rs}{\ensuremath{R_\odot}}

\newcommand{\kms}{km s$^{-1}$}

\authorrunning{Decressin et al.}
\titlerunning{Secular Hydrodynamical Processes in Rotating Main Sequence Stars}
\begin{document}
      \title{Diagnoses to unravel secular hydrodynamical processes\\ in  rotating main sequence stars} %

   \author{T. Decressin
          \inst{1,2},
          S. Mathis \inst{3,8},
          A. Palacios \inst{4},
          L. Siess \inst{5},
          S. Talon\inst{6},
          C. Charbonnel \inst{1,7},
          J.-P. Zahn \inst{8}
          }

   \offprints{S. Mathis or T. Decressin}

   \institute{Geneva Observatory, University of Geneva, chemin des
     Maillettes 51, CH-1290 Sauverny, Switzerland
                \and
                Argelander Institute for Astronomy (AIfA), Auf dem
                H\"ugel 71, D-53121 Bonn, Germany\\
                \email{decressin@astro.uni-bonn.de}
                \and
               Laboratoire AIM, CEA/DSM-CNRS-Universit\'e Paris Diderot, IRFU/SAp Centre de Saclay, F-91191 Gif-sur-Yvette, France\\
                \email{stephane.mathis@cea.fr}
               \and
               GRAAL, Universit\'e Montpellier II, CNRS, Place
               E. Bataillon, F-34095 Montpellier Cedex 05, France
               \and
               IAA-ULB, Universit\'e Libre de Bruxelles, Boulevard du
               Triomphe, CP 26, B-1050 Bruxelles, Belgium
               \and
               R\'eseau Qu\'eb\'ecois de Calcul de Haute Performance,
               Universit\'e de Montr\'eal (DGTIC), CP 6128, succ. Centre-ville,
               Montr\'eal H3C 3J7, Canada
               \and
               LATT, CNRS UMR 5572, Universit\'e de Toulouse, 14 avenue Edouard
               Belin, F-31400 Toulouse Cedex 04, France
               \and
              LUTH, Observatoire de Paris-CNRS-Universit\'e Paris-Diderot, Place Jules Janssen, F-92195 Meudon, France\\
             }

   \date{Received; accepted }

   \abstract {Recent progresses and constraints brought by helio\- and
     astero\-seismology call for a better description of stellar
       interiors and an accurate description of rotation-driven mechanisms
       in stars.}%
   {We present a detailed analysis of the main physical
     processes responsible for the transport of angular momentum and
     chemical species in the radiative regions of rotating stars. We focus
     on cases where meridional circulation and shear-induced turbulence
     only are included in the simulations (i.e., no internal gravity waves
     nor magnetic fields). We put special emphasis on analyzing the angular
     momentum transport loop and on identifying the contribution of each
     physical process involved.}%
   {We develop a variety of diagnostic tools designed to help
     disentangle the role of the various transport mechanisms. Our analysis
     is based on a 2-D representation of the secular hydrodynamics, which
     is treated using expansions in spherical harmonics. By taking
     appropriate horizontal averages, the problem reduces to one dimension,
     making it implementable in a 1D stellar evolution code while
     preserving the advective character of angular momentum transport. We
     present a full reconstruction of the meridional circulation and of the
     associated fluctuations of temperature and mean molecular weight
     along with diagnosis for the transport of angular momentum, heat
     and chemicals. In the present paper these tools are used to validate 
     the analysis of two main sequence stellar models of 1.5 and
       20~\Ms{} for which the hydrodynamics has been previously extensively
       studied in the literature.}%
   {We obtain a clear visualization and a precise estimation of the
       different terms entering the angular momentum and heat transport
       equations in radiative zones of rotating stars. This enables us to
       corroborate the main results obtained over the past decade by Zahn,
       Maeder, and collaborators concerning the secular hydrodynamics of
       such objects. We focus on the meridional circulation driven by
       angular momentum losses and structural readjustements. We confirm
       quantitatively for the first time through detailed computations and
       separation of the various components that the advection of entropy
       by this circulation is very well balanced by the barotropic effects
       and the thermal relaxation during most of the main sequence
       evolution. This enables us to derive simplifications for the thermal
       relaxation on this phase. The meridional currents in turn advect
       heat and generate temperature fluctuations that induce differential
       rotation through thermal wind thus closing the transport loop. We
       plan to make use of our refined diagnosis tools in forthcoming
       studies of secular (magneto-)hydrodynamics of stars at various
       evolutionary stages.}%
   {}

               
\maketitle
%
\section{The impact of rotation on stellar evolution}

Rotation, and more precisely differential rotation, has a major impact on
the internal dynamics of stars, in several ways. It 
induces large-scale circulations both in radiative and convective
zones that advect simultaneously angular momentum, nuclides and magnetic
fields (\citealp{E25,V25,S50,M53,busse82,zahn92,Talon97,TZMM97,MZ98,MM00,Garaud02b,PA03,PCTS06,MZ04,MZ05,Rieutord06,ER07,MPZ07}).

When the star rotates differentially, various instabilities develop
(secular and dynamical shear instabilities, baroclinic and multidiffusive
instabilities) that generate hydrodynamical turbulence in radiative
zones, in addition to these circulations. Just as in the terrestrial
atmosphere and in laboratory experiments, this turbulence acts so as to
reduce its cause, namely the gradients of angular velocity and of chemical
composition. This explains why its effect may be described as a diffusion
process \citep[][and references
therein]{TZ97,Maeder03,MPZ04,talon07}.

Rotating stars have an equator that is cooler than the poles,
which has an important effect on radiatively driven stellar winds and hence
on the loss of mass and angular momentum \citep{M99}. If rotation is large
enough, the stars can even reach the break-up limit, when centrifugal forces
balances the gravity, and matter is ejected through an equatorial
mechanical wind which create a circumstellar disk \citep{MaMe00,M07,E08}.

Rotation also has a strong impact on the stellar magnetism. 
In radiative zones, rotation interacts with the magnetic field and is
able to trigger magnetohydrodynamical instabilities,
that could play a role in the transport of angular momentum and nuclides
\citep{CMG93,Garaud02a,Spruit99,Spruit02,Menou04,MM04,Egg05,BS05,MZ05,B06,BZ06,ZBM07}.

Furthermore, internal gravity waves and gravito-inertial waves that are
excited at the edge of the convection zones may also contribute to the
transport of angular momentum.  They propagate inside radiative zones and
they extract or deposit angular momentum in the region where they are
damped, thus modifying the angular velocity profile and the vertical
distribution of chemicals
\citep{Press81,Schatzman93,TKZ02,TC03,TC04,TC05,TC08,CT05,RG05,Pantillon07,Mathis08}.

All these effects significantly modify the evolution of rotating stars.
They affect their surface velocities and chemical abundances, and change
their paths across the colour-magnitude diagram \citep{MM00b,M09}.  They
also modify the internal structure of stars in a way that we will soon be
able to test thanks to asteroseismology over a broader range of
evolutionary phases.

In this paper we present a set of diagnosis tools adapted to the analysis
of stellar evolution with rotation, and use them to compare the efficiency
of different transport processes.  The present study focuses on ``type I
rotational transport'', where magnetic fields and waves are not accounted
for, and where the angular momentum and the nuclides are both transported
exclusively by meridional circulation and shear-induced turbulence.  These
diagnosis and tools rely on a specific expansion of the equations for the
transport of angular momentum, heat and chemicals that are briefly recalled
in \S~2. In \S~3, we validate our method by applying our tools to two
specific stellar models for which the hydrodynamics has already been
extensively studied in the literature. Conclusions and perspectives are
presented in \S~4.

\section{Modelling the secular processes}\label{sec:form}

\subsection{Scale separation}

To simulate in full detail the dynamical processes in a star, would
require to include length-scales and time-scales spanning many orders of
magnitude. This is clearly not feasible, even with the most powerful
computers. Either one chooses to describe what occurs on a dynamical
time-scale, such as a convective turnover time, or one focuses on the long
time evolution, as we shall here, where the typical time is either the
Kelvin-Helmholtz time or that characterizing the dominant nuclear
reactions. The same is true for the length-scales, at least in the vertical
direction, where we shall take the resolution which represents adequately
the steepest gradients that develop during the evolution.

\begin{figure}[tp]
\centering
\includegraphics[width=0.53\textwidth]{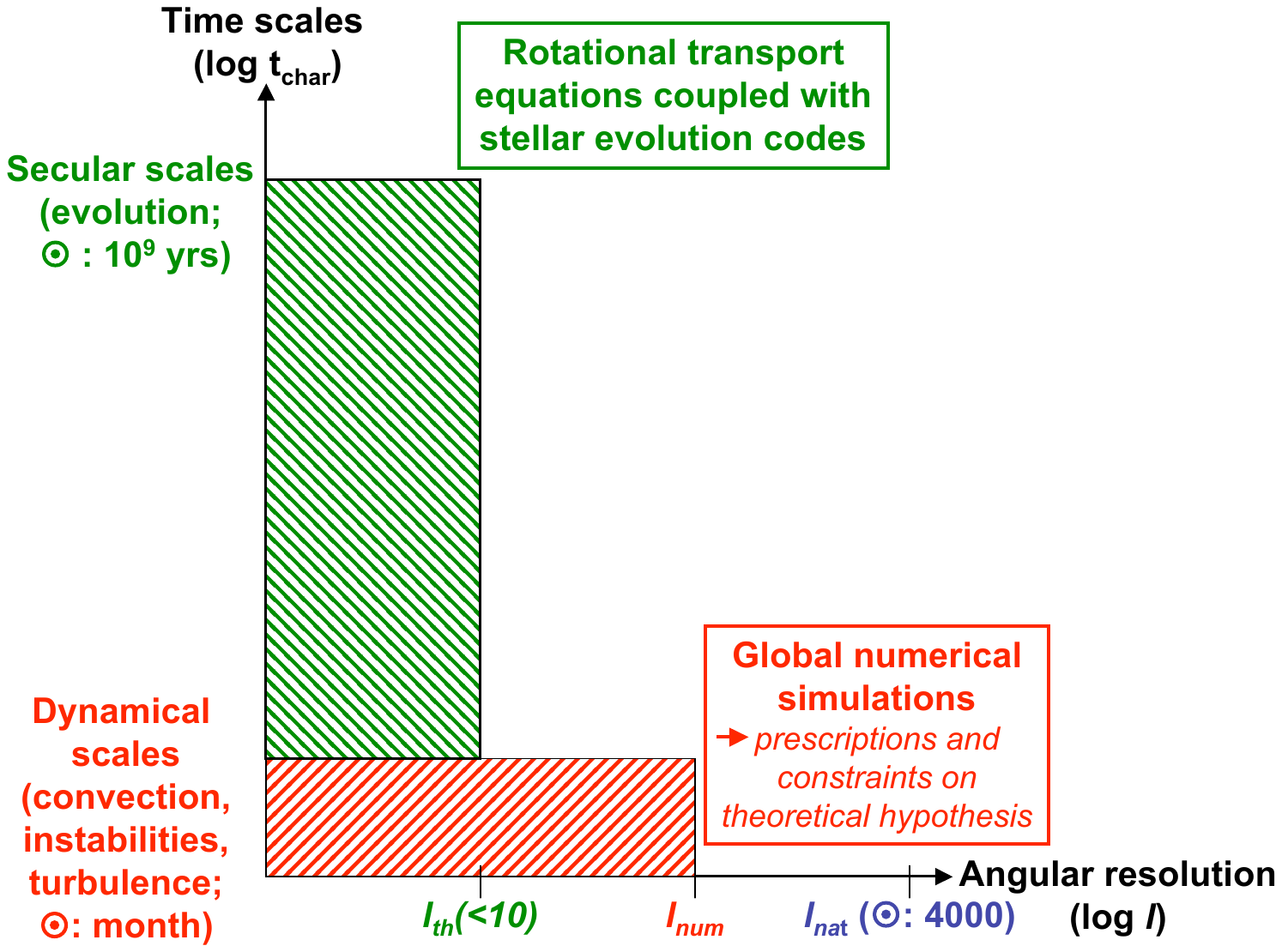}
\caption{Modelling strategy to study dynamical stellar evolution. The
  diagram presents timescales of the typical physical processes as a
  function of the angular resolution necessary to properly describe these
  processes. The angular resolution is expressed in terms of the $l$ index
  of the spherical harmonics $Y_{l,m}\left(\theta,\phi\right)$. $l_{\rm
  num} \simeq 600 $ indicates the maximum angular resolution (in term of spherical
  harmonics nodes) presently achieved in global numerical
  simulations.}
\label{diagsecular}
\end{figure}

The situation is somewhat different in the horizontal direction -- i.e. in
latitude, since we consider here only the axisymmetric case. Stellar
radiative zones are stably stratified regions, and the buoyancy, which is
the restoring force, acts to inhibit turbulent motions in the vertical
direction. This leads to a strongly anisotropic turbulent transport that is
more efficient in the horizontal direction (along isobars) than in the
vertical one. As a result, the horizontal gradients of all scalar fields
(temperature, angular velocity\ldots) are much smaller than their
vertical gradients, which allows their expansion in a few spherical
harmonics. This scale separation, both in space and in time, is illustrated in Fig.~\ref{diagsecular}.

Considering the axisymmetric case, each scalar field is thus written as the
sum of its horizontal average on an isobar, and its associated fluctuation,
which is expanded the Legendre polynomials basis $P_\ell(\cos \theta)$, up
to some maximum order $\ell_{th}$. For simplicity's sake, we shall present
here the results keeping a single scale $\ell=2$, that is dominant.
Let us note that there are circumstances where we have to include
many more components, for instance when dealing with magnetic fields.

The unresolved scales intervene in the turbulent transport, for which a
prescription is applied that is derived, whenever possible, from laboratory
experiments or numerical simulations, and if not through phenomenological
considerations.

\subsection{Linearization and expansion in spherical harmonics}

Let us briefly recall how these expansions are performed in practice; for
more details we refer the reader to \citet{MZ04} and \citet{M09}.  
We begin with the macroscopic velocity field, which is split in 3 components:
\begin{equation}
\vec V=\underbrace{r\sin\theta \,\Omega\left(r,\theta\right){\widehat {\bf e}}_{\varphi}}_{1}+\underbrace{\dot r \,{\widehat {\bf e}}_{r}}_{2}+\underbrace{\vec{\mathcal U}_{M}\left(r,\theta\right)}_{3} \, .
\end{equation}
Term 1 represents the azimuthal velocity field associated with the
differential rotation. Term 2 corresponds to the radial Lagrangian velocity
due to the structural readjustments of the star during its evolution. Term
3 is the meridional circulation velocity field. $\widehat{\bf e}_{k}$ where
$k=\left\{r,\theta,\varphi\right\}$ are the unit vectors in the radial,
latitudinal and azimuthal directions.

In radiative zones, because of the strong stratification, the turbulent transport
is highly anisotropic, which leads to a ``shellular'' rotation profile
\citep{zahn92}. The
angular velocity thus varies little on an isobar, allowing to expand it
as
\begin{equation}
\Omega(r, \theta) = \overline\Omega(r)+\Omega_2(r)\left[P_2(\cos
  \theta)-{1/ 5}\right] , 
\end{equation}
\noindent where
\vskip-6pt
\begin{equation}
\overline\Omega\left(r\right)=\frac{\int_{0}^{\pi}\sin^3\theta\,\Omega\left(r,\theta\right)\,{\rm d}\theta}{\int_{0}^{\pi}\sin^{3}\theta\,{\rm d}\theta}\, .  
\end{equation}

As usual, the meridional velocity field is projected on spherical
functions, of which we show here only that of lowest order:
\begin{equation}
\vec{\mathcal
  U}_{M}=U_{2}\left(r\right)P_{2}\left(\cos\theta\right){\widehat {\bf
    e}}_{r}+V_{2}\left(r\right) \frac{{\rm
    d}P_{2}\left(\cos\theta\right)}{{\rm d}\theta}{\widehat {\bf
    e}}_{\theta}\, , 
\label{eq:Um}
\end{equation}
where $P_2$ is the second order Legendre polynomial. The anelastic
approximation is adopted, filtering out the acoustic waves, which is amply
justified for this slow meridional circulation. Therefore, the continuity
equation reduces to $\vec\nabla\cdot\left(\overline\rho\vec{\mathcal
    U}_{M}\right)=0$, which leads to the following relation between the
latitudinal ($V_2$) and the radial ($U_2$) amplitude functions:
\begin{equation}
V_2=\frac{1}{6\overline\rho r}\partial_{r}\left(\overline\rho r^2 U_{2}\right).
\label{eq:V2}
\end{equation}
Moreover, since $\overline\rho\vec{\mathcal U}_{M}$ is divergenceless, it may be expressed using a stream function ($\xi_{\rm M}\left(r,\theta\right)$) as in Spiegel \& Zahn (1992), where
\begin{equation}
\overline\rho \vec {\mathcal U}_{\rm M}=-\frac{1}{r\sin\theta}\vec\nabla\xi\times\widehat{\bf e}_{\varphi}
\label{SF}
\end{equation}
that leads to
\begin{equation}
\overline\rho{\mathcal U}_{\rm M;r}=-\frac{1}{r^2\sin\theta}\partial_{\theta}\,\xi_{\rm M}\hbox{ }\hbox{ }\hbox{and}\hbox{ }\hbox{ }\overline\rho{\mathcal U}_{\rm M;\theta}=\frac{1}{r\sin\theta}\partial_{r}\,\xi_{\rm M} 
\end{equation}
with
\vskip-18pt
\begin{equation}
\xi_{\rm M}=\frac{1}{2}\overline\rho r^2 U_2 \left(\cos^3\theta-\cos\theta\right).
\label{eqUloop}
\end{equation}
${\overline\rho}{\vec {\mathcal U}}_{\rm M}$, and thus ${\vec {\mathcal U}}_{\rm M}$, are then tangent to the iso-$\xi$ lines since $\vec{\mathcal U}_{\rm M}\cdot{\vec\nabla}\xi=0$ ({\it c.f.} Eq. \ref{SF}).

\medskip

Finally, all other variables are expanded as shown here for the temperature
and the mean molecular weight:
\begin{equation}
T\left(r,\theta\right)\!=\!{\overline T}\left(r\right)\!+\!\delta T\left(r,\theta\right)\hbox{ }\hbox{ }\hbox{with}\hbox{ }\hbox{ }\delta T\left(r,\theta\right)\!=\!\left[\Psi_{2}\left(r\right){\overline T}\right]P_{2}\left(\cos\theta\right)
\label{eq:deltaT}
\end{equation}
\vskip-6pt
\noindent and
\vskip-12pt
\begin{equation}
\mu\left(r,\theta\right)\!=\!{\overline \mu}\left(r\right)\!+\!\delta \mu\left(r,\theta\right)
\hbox{ }\hbox{ }\hbox{with}\hbox{ }\hbox{ }\delta
\mu\left(r,\theta\right)\!=\!\left[\Lambda_{2}\left(r\right)\overline{\mu}\right]P_{2}\left(\cos\theta\right);
\label{eq:deltamu}
\end{equation}
${\overline T}$ and ${\overline \mu}$ are the horizontal averages, ${\delta
  T}$ and $\delta \mu$ being their fluctuations and $\Psi_{2}$ and
$\Lambda_{2}$ their relative fluctuations on the isobar.

\subsection{The transport equations}

Using these notations, the transport equations for angular momentum 
and chemicals may thus be expanded in the following way.

\subsubsection{Transport of angular momentum and thermal-wind}

We first consider the azimuthal projection of the momentum equation. By
averaging it over an isobar, we obtain the following advection-diffusion
equation for the mean angular momentum :
\begin{equation}
{\overline\rho}\frac{{\rm d}}{{\rm d}t}\left(r^2\overline{\Omega}\right)=\underbrace{\frac{1}{5r^2}\partial_{r}\left({\overline\rho} r^4 \overline{\Omega}U_{2}\right)}_{1}+
\underbrace{\frac{1}{r^2}\partial_{r}\left({\overline\rho}\nu_{v}r^4\partial_{r}\overline{\Omega}\right)}_{2}.
\label{AM}
\end{equation}
Term 1 represents the transport of angular momentum by the meridional
circulation; note that the advective character of that
transport is preserved.  The diffusion term 2 is associated with the action of the
shear-induced turbulence, where $\nu_v$ is the turbulent viscosity in the
vertical direction (see below Eq.~\ref{eq:dv-mu}). The Lagrangian time
derivative ${\rm d}/{{\rm d}t}$ is defined as ${\rm d}/{\rm
  d}t=\partial_{t}+{\dot r}\partial_{r}$.
  
This equation shows the relation between the meridional circulation and the
transport of angular momentum. In the asymptotic regime, the left-hand side
term is zero and the transport of angular momentum by meridional
circulation is exactly balanced by that through shear turbulence. In the
limit of vanishing turbulent viscosity, the rotation profile would adjust
so that no meridional currents appear \citep{busse82}. In reality because
of the loss of angular momentum by the wind, and/or its redistribution by
structural changes, the left-hand side term is non zero.

By integrating Eq.(\ref{AM}) over a surface of radius $r$, it can be recast in an equation for the fluxes:
\begin{equation}
\Gamma(m) = -F_{\rm tot} = - F_{\rm MC}\left(r\right) - F_{\rm S}\left(r\right),
\label{fluxAM}
\end{equation}
where 
\begin{equation}
F_{\rm MC}\left(r\right)=-\frac{1}{5}\overline\rho r^4{\overline\Omega}U_{2}
\label{eq:Fmc}
\end{equation}
is the flux transported by the meridional circulation, and
\begin{equation}
F_{\rm S}\left(r\right)= - \overline\rho r^4 \nu_{v}\partial_{r}{\overline\Omega}
\label{eq:FS}
\end{equation}
is that carried by the vertical shear induced turbulence. The term $\Gamma(m)$
represents the loss or gain of angular momentum inside the isobar enclosing
the mass $m\left(r\right)$; its derivation is given in Appendix
\ref{ap:A1}:
\begin{equation}
\Gamma\left(m\right)=\frac{1}{4\pi}\frac{\rm d}{{\rm d}t}\left[\int_{0}^{m\left(r\right)}{r'}^2\overline{\Omega}\, {\rm d}m'\right] .
\label{eq:gamma}
\end{equation}
From Eq. (\ref{fluxAM}) it is also possible to extract the radial component
of the meridional circulation velocity:
\begin{equation}
U_{2}=U_\Gamma+U_V=\frac{5}{\overline\rho r^4 \overline{\Omega}}\left[\Gamma\left(m\right)-\overline\rho\nu_{v}r^4\partial_{r}\overline{\Omega}\right] .
\label{eq:ucirc}
\end{equation}
Finally, by taking the curl of the momentum equation and again filtering
out the fast timescales, we obtain the so-called thermal wind equation:
\begin{equation}
\frac{1}{3}\,r\,\partial_{r}{\overline\Omega}^2=\frac{\overline g}{r}\left(\varphi\Lambda_{2}-\delta\Psi_{2}\right)
\label{ThermalWind}
\end{equation}
where $\overline g$ is the horizontal average of the effective
gravity\footnote{In a rotating star, the centrifugal force diminishes the
  local gravity, and the {\em effective} gravity $\overline g$ is then
  given by horizontally averaging the hydrostatic equilibrium : $ \vec{g} =
  \vec\nabla\Phi + \frac{1}{2} \Omega^2 \nabla( r \sin \theta)^2$, where
  $\Phi$ is the gravitational potential.}. The coefficients
$\delta=-\left(\partial\ln \rho/\partial\ln T\right)_{P,\mu}$ and
$\varphi=\left(\partial\ln \rho/\partial\ln \mu\right)_{P,T}$ are
introduced by the generalised equation of state (\citealp{KW90}; see
\citealp{MZ98} or \citealp{MZ04} more details).  The right-hand side term
is the baroclinic torque $\frac{1}{\rho}\vec\nabla \rho \times \vec g$.

\subsubsection{Thermal relaxation}

Likewise, by expanding the energy conservation equation over spherical
functions, we establish the following equation for the temperature
perturbation represented here by $\Psi_2$ (Eq. 101 in \citealp{MZ04},
see \citealp{zahn92}):
\begin{equation}
\underbrace{{\overline \rho}{\overline T}C_{p}\frac{{\rm d}\Psi_{2}}{{\rm d}t}}_{{\overline\rho}{\overline T}\partial_{t}{\widetilde S}}=\underbrace{-{\overline \rho}{\overline T}C_{p}\frac{N_{T}^{2}}{{\overline g}\delta}U_{2}}_{{\overline\rho}{\overline T}U_{r}\partial_{r}{\overline S}}+\underbrace{{\overline \rho}\frac{L}{M}{\mathcal T}_{2,{\mathcal B}}+{\overline \rho}\frac{L}{M}{\mathcal T}_{2,{\rm Th}}}_{\vec\nabla\cdot\left(\chi\vec\nabla T\right)-\vec\nabla\cdot\vec F_{H}}+\underbrace{{\overline \rho}\frac{L}{M}{\mathcal T}_{2,{\rm N-G}}}_{\rho\varepsilon}
\label{eq:Mer}
\end{equation}
where $N_{T}$, the buoyancy frequency linked with the entropy stratification is given by: $N_{T}^2=({\overline g}\delta/{H_{P}})\left(\nabla_{\rm ad}-\nabla\right)$ with the usual notations for the temperature gradients: $\nabla=\partial\ln{\overline T}/\partial \ln P$.
This is again an advection/diffusion equation, where the advective term
plays the role of a heat source or sink, since the meridional circulation
is caused by the transport of angular momentum, with the other terms
describing essentially the thermal relaxation due to radiative damping.
$\vec\nabla\cdot{\vec F}_{H}$ is the entropy flux carried by the horizontal turbulence.

For diagnostic purpose (and also for historical reasons; see the discussion
in \citealt{zahn92} after Eq.~(3.32))  we split the
diffusion term in two pieces, which expressions are given in Appendix
\ref{ap:A2}:
\begin{itemize}
\item ${\mathcal T}_{2,{\mathcal B}}$ corresponds to the divergence of the
  radiative flux associated to the deformation of the isobar induced by the
  perturbing force, which is here the centrifugal acceleration. This term
  subsists in the case of solid body rotation; for this reason we called it
  the {\sl barotropic} term following \citet{zahn92}.
\item ${\mathcal T}_{2,{\rm Th}}$ contains the highest derivatives of the
  diffusion operator and according to the thermal wind equation
  (\ref{ThermalWind}) it vanishes in an homogeneous star when the rotation
  is uniform. It is referred to as the thermic or {\sl baroclinic} term
  \citep{zahn92}.
\end{itemize}
The last term ${\mathcal T}_{2,\rm N-G}$ is associated with the nuclear
reactions and radial structural readjustments during its evolution and
usually plays a negligible role.

Eq. (\ref{eq:Mer}) can finally be written in the following form
\begin{equation}
{\tilde{\mathcal T}}_{\rm N-S}={\tilde{\mathcal T}}_{\rm Adv}+{\tilde{\mathcal T}}_{\mathcal B}+{\tilde{\mathcal T}}_{\rm Th}+{\tilde{\mathcal T}}_{\rm N-G},
\label{RelaxThEq}
\end{equation}
where
\begin{equation}
{\tilde{\mathcal T}}_{\rm NS}={\overline \rho}{\overline T}C_{p}\frac{{\rm d}\Psi_{2}}{{\rm d}t},
\end{equation}
\begin{equation}
{\tilde{\mathcal T}}_{\rm Adv}=-{\overline \rho}{\overline T}C_{p}\frac{N_T^2}{{\overline g}\delta}U_2,
\end{equation}
\begin{equation}
{\tilde{\mathcal T}}_{\mathcal B}={\overline\rho}\frac{L}{M}{\mathcal T}_{2,\mathcal B},
\end{equation}
\begin{equation}
{\tilde{\mathcal T}}_{\rm Th}={\overline\rho}\frac{L}{M}{\mathcal T}_{2,{\rm Th}},
\end{equation}
\begin{equation}
{\tilde{\mathcal T}}_{\rm N-G}={\overline\rho}\frac{L}{M}{\mathcal T}_{2,{\rm N-G}}.
\end{equation}
This will be used in \S 3.4. in our diagnosis tools.

\begin{figure*}[tbp]
\begin{center}
\includegraphics[width=0.375\textwidth]{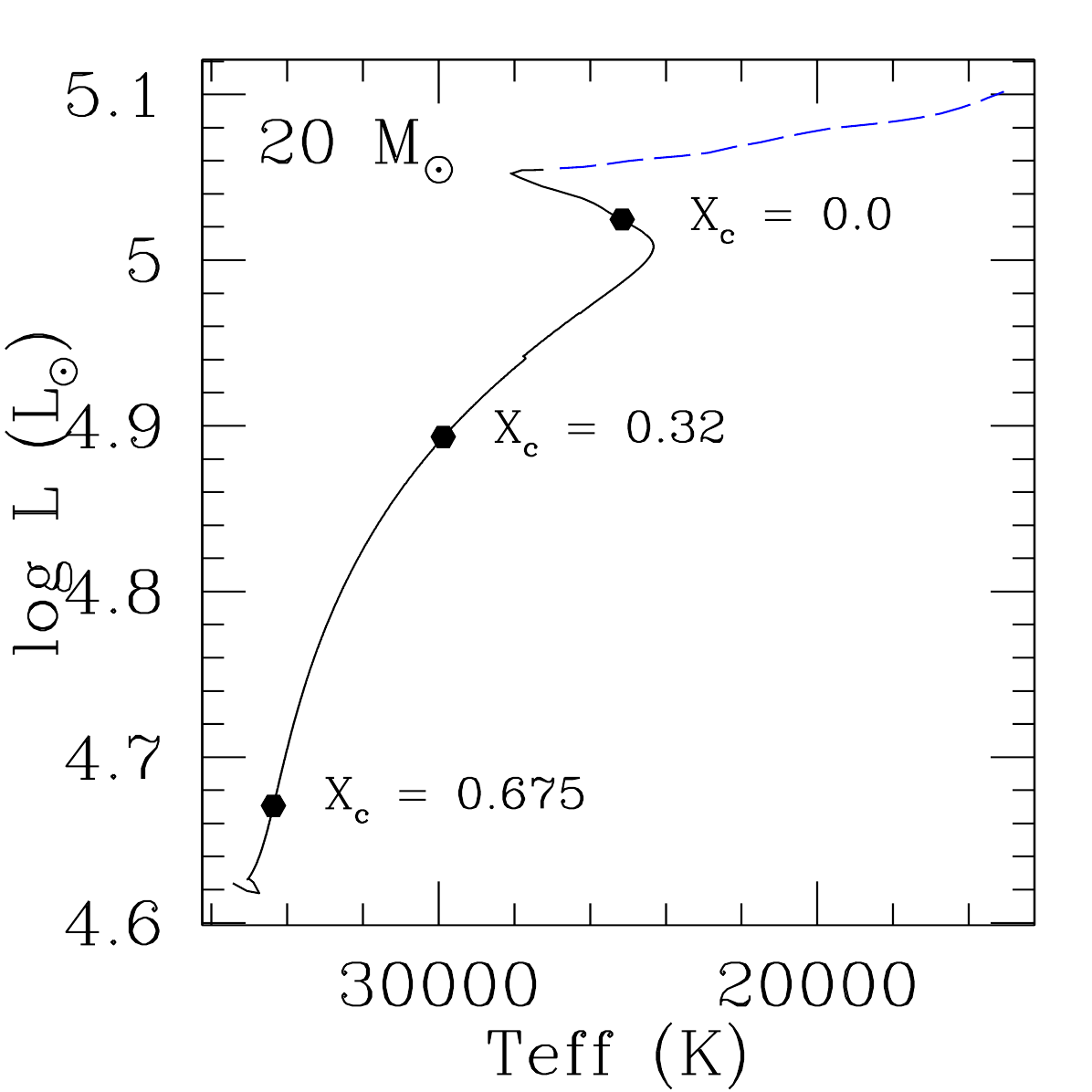}
\includegraphics[width=0.375\textwidth]{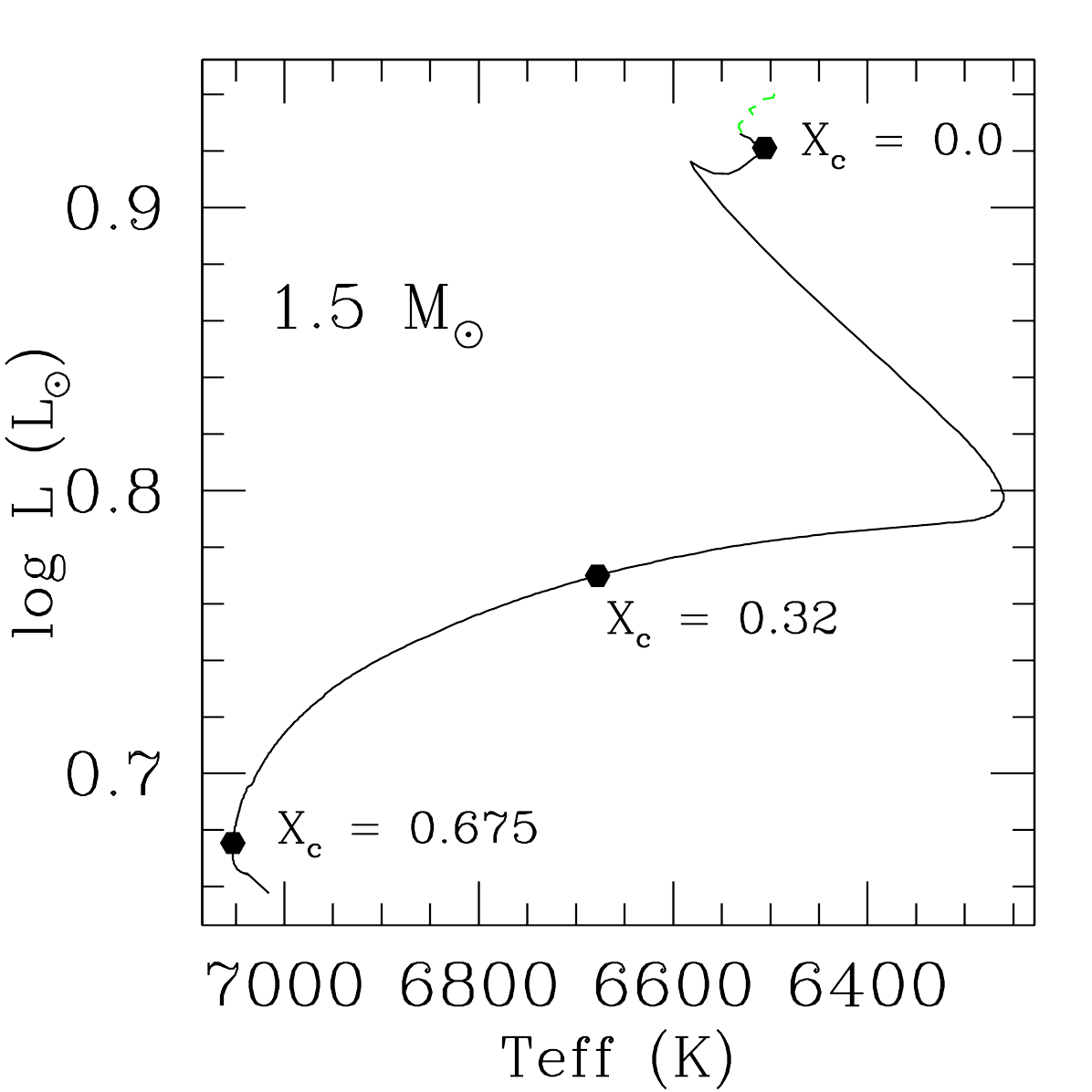}
\caption{Evolutionary path in the HR diagram of a 20~\Ms{} (left panel) and
  1.5~\Ms{} (right panel) star with solar composition. The solid line
  represents the main sequence. The dots on each plot indicate specific
  locations at which a detailed analysis of the structure is presented.}
\label{fig:HR}
\end{center}
\end{figure*}

\subsubsection{Transport of nuclides}

The expansion of the transport equation for the nuclides on an isobar leads to an
equation for the evolution of the mass fraction of each considered
chemical \citep[see also][]{MM00}:
\begin{eqnarray}
  \left(\frac{{\rm d} X_i}{{\rm d}t}\right)_{M_r}= \frac{\partial}{\partial
  M_r}\left[(4\pi
  r^2\rho)^2\left(D_v+D_{\rm eff}\right) \frac{\partial X_i}{\partial M_r}\right] + \left(\frac{{\rm d}X_i}{{\rm d}t}\right)_{\rm nucl},
\label{Cmoy}
\end{eqnarray}
where $d M_{r}=4\pi{\overline\rho} r^2 d r$ and $D_v$ is the vertical component of the turbulent diffusivity
  (see Eq.~\ref{eq:dv-mu} below). The
strong horizontal turbulence leads to the erosion of the advective
transport, which can then be described as a diffusive process \citep{CZ92}
with the following effective diffusion coefficient
\begin{equation}
D_{\rm eff}=\frac{\left(r U_2\right)^2}{30 D_h},
\label{DEF_DEFF}
\end{equation}
where $D_h$ the horizontal component of the turbulent diffusivity (see
Eqs.~\ref{nuhg} and \ref{nuhz}).

The second term on the right-hand
side of Eq.~(\ref{Cmoy}) corresponds to the temporal mass fraction evolution of the $i^{th}$
nuclide due to nuclear burning.

Equation (\ref{Cmoy}) is complemented by an equation for the time evolution of the
relative fluctuation of the mean molecular weight, expressed here in terms
of $\Lambda_2$ :
\begin{equation}
\frac{{\rm d}\Lambda_{2}}{{\rm d}t} - \frac{{\rm d} \ln {\overline
    \mu}}{{\rm d}t} \Lambda_{2} = \frac{N_{\mu}^{2}}{{\overline
    g}\varphi}U_{2}-\frac{6}{r^2}D_h\Lambda_{2}\, , 
\label{Lambda}
\end{equation}
 where $N_{\mu}$, the chemical part of the  Brunt-V\"ais\"al\"a
  frequency, is given by $N_{\mu}^2=\left({\overline
      g}\varphi/H_{P}\right)\nabla_{\mu}$ with $\nabla_{\mu}={\partial \ln 
    \overline{\mu}}/{\partial \ln P}$. 

\subsubsection{Turbulence modelling}\label{turb-mod}

The details of turbulence modelling have already been extensively discussed
in previous papers \citep{TZMM97,PCTS06}, so that  we
will just recall  here the expressions we have chosen.

For the vertical turbulent diffusion coefficients ($D_v \simeq \nu_v$) we
use the expression by \citet{TZ97}: 
\begin{equation}
D_v = { Ri_{\rm c} \over N^2_{T} /(K_{T} + D_{h}) + N^{2}_{\mu}/ D_{h}}
\left(r\partial_{r}\overline\Omega\right)^2\,; 
\label{eq:dv-mu}
\end{equation}
where $Ri_{\rm c} = 1/6$ is the adopted value for the critical Richardson
number and $K_T$ the thermal diffusivity.

 For the horizontal turbulent viscosity we have used two different prescriptions:
\begin{equation}
\nu_h=r\sqrt{\left[Cr\overline\Omega \vert 2 V_{2}-\alpha U_{2}\vert\right]}
\quad \hbox{where} \; \alpha={1\over 2}\,
\frac{{\partial}\ln(r^2\overline\Omega)}{{\partial}\ln r} , 
\label{nuhg}
\end{equation}
  after \citet{MPZ04}, and
 \begin{equation}
 \nu_h=\frac{r}{C_{H}}\vert2 V_{2}-\alpha U_2\vert
 \label{nuhz}
 \end{equation}
 after \citet{zahn92}. $C = 1.6\times10^{-6}$, $C_{H}$ is a parameter of order
 unity, and $V_2$ is given by Eq.~(\ref{eq:V2}). As for the vertical
turbulent diffusion coefficient, we assume that $D_h \simeq \nu_h$. 

We stress that the advective character of the transport of angular momentum
by the meridional circulation makes the interpretation of the whole
hydrodynamics more complex than when the diffusive approximation is used
\citep[e.g.,][]{Endal78,Pinsonneault89,Heger00}.
This is one of the reasons that incited us to develop the tools that we
describe in the following section.

\begin{figure*}[tbp]
\begin{center}
\includegraphics[width=0.375\textwidth]{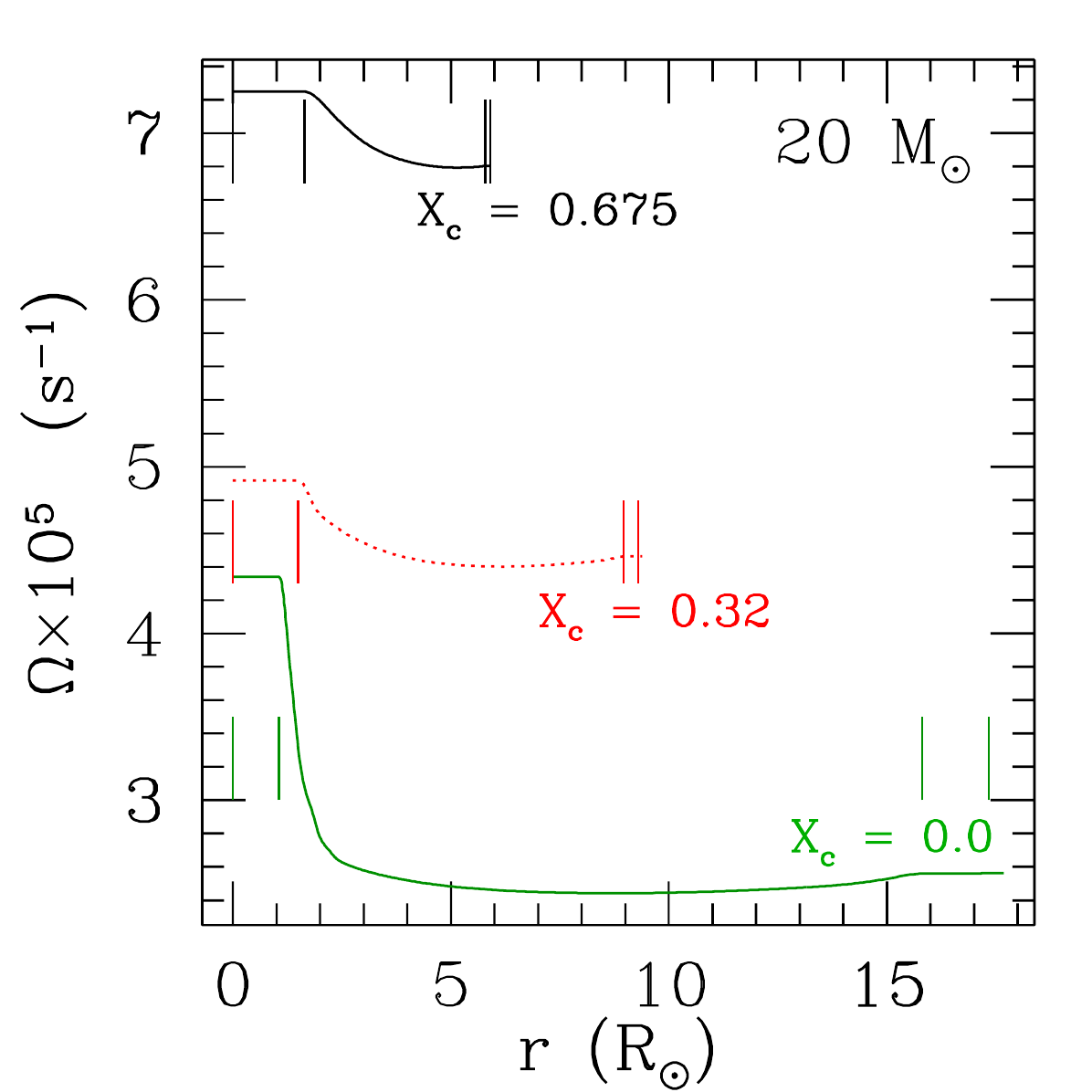}
\includegraphics[width=0.375\textwidth]{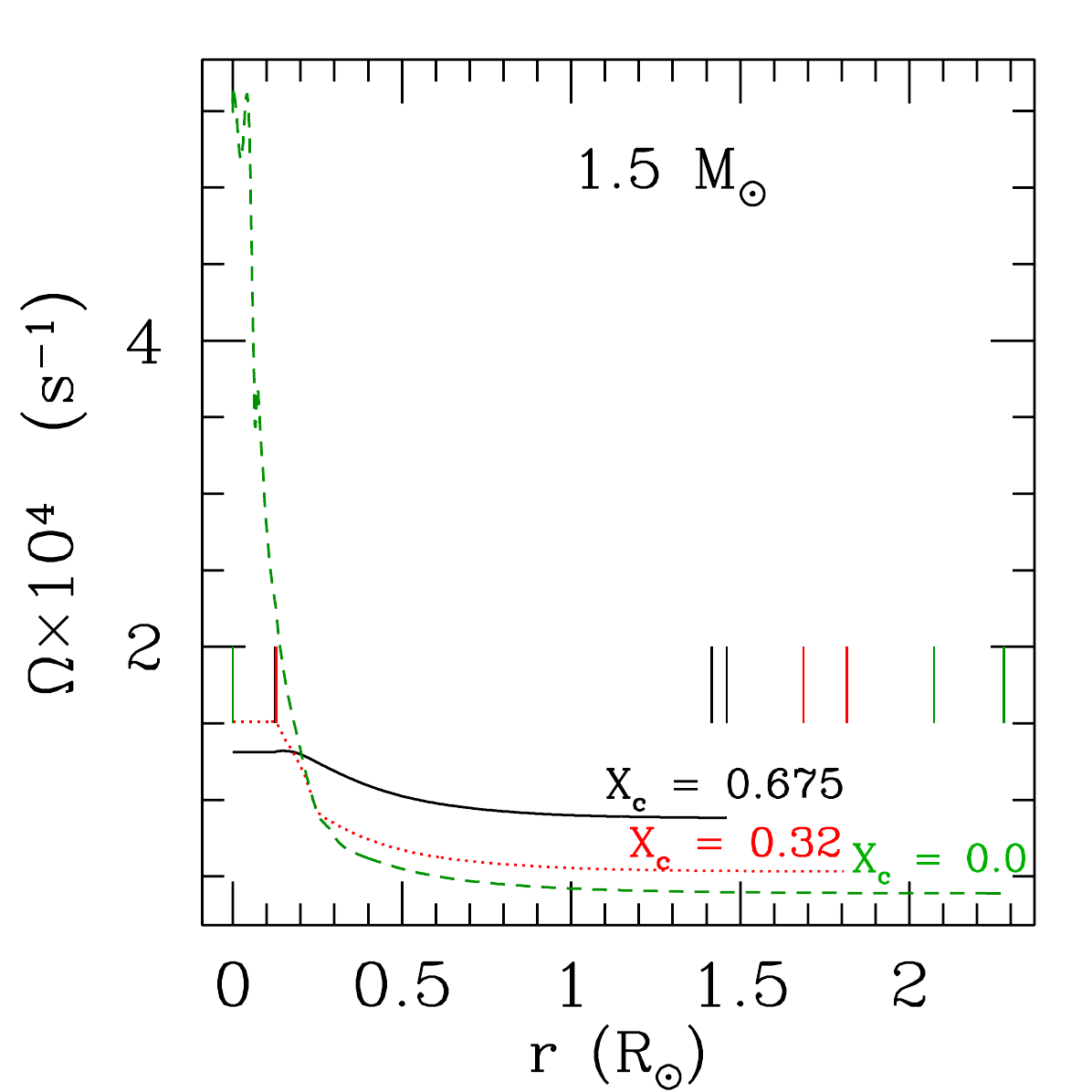}
\caption{Angular velocity profile inside the 20~\Ms{} ({\em left panel})
  and 1.5~\Ms{} ({\em right panel}) model as a function of the (expanding)
  radius, at three different evolutionary stages on the main sequence, as
  indicated by the central hydrogen content. The short vertical lines
  indicate the location of the convection zones. In the 1.5~\Ms{} model, at
  the end of the main sequence ($X_\text{c}$ = 0), the convective core
  has disappeared.}
\label{fig:omega}
\end{center}
\end{figure*}

\section{Numerical simulation of secular transport}

In this section, we apply the formalism described in \S~\ref{sec:form} to
the evolution of rotating stars with initial masses of 20 and
1.5~\Ms{} and metallicity close to solar value ($Z = 0.017$ and 0.02
  for the 20 and the 1.5~\Ms{} respectively). In the
  Hertzsprung-Russell diagram shown in Fig.~\ref{fig:HR}, the black dots
  indicate the evolutionary points where a snapshot on the internal
  structure will be taken and analysed in details. These points are
associated with central hydrogen mass fractions $X_\text{c} =
0.675,~0.32,~0.00$ as indicated on the tracks.

A number of results concerning the hydrodynamics of such rotating
  models have already been published
\citep{Talon97,TC98,MM00,PA03,MPZ04,PCTS06,DMCPE07}, and we will rely on
this experience to prove the relevance of the diagnosis tools that we have
developed. The models presented here were computed with the stellar
evolution code STAREVOL V2.90, and the reader is referred to
\citet{Siess00}, \citet{PCTS06} and \citet{Siess06} for a detailed
description of the input physics. We shall just recall the main
characteristics and parameters used for the modeling.

The reference solar composition we use follows \citet{GNS96}.
The standard mixing length theory is used to model the temperature gradient
in the convection zones and the parameter $\alpha_{\rm MLT} = 1.75$. The
atmosphere is treated in the gray approximation, and integrated up to an
optical depth $\tau \approx 5 \times 10^{-3}$.

In both simulations mass loss is included from the Zero Age Main Sequence
on. We use the \citet{Reimers75} prescription for the low-mass star, with a
parameter $\eta_{\rm R} = 0.5$ \citep[see also][]{PCTS06} and the
prescriptions by \citet{Vink00} for the massive star. Rotation effects
  on mass loss are accounted for following \citet{MM01}. Angular momentum
losses associated with mass loss are also accounted for, but their
  anisotropy \citep{Maeder02} is not included in the computations.

Concerning the transport of angular momentum and nuclides, we use
Eq.~(\ref{nuhg}) as the prescription for $D_h$ for the 20~\Ms{} model.
For the 1.5~\Ms{} model, we use Eq.~(\ref{nuhz}) as in \citet{PA03} 
who showed that it yields to a very good
agreement with the surface velocity and the light nuclides (Li, Be) surface
abundances that are observed in open clusters such as the Hyades.  The
expression for the vertical turbulent diffusion coefficient is given by
Eq.~(\ref{eq:dv-mu}) in both models. For the transport of nuclides, we do
not account for atomic diffusion.

In our framework, the evolution of the angular velocity profile and of the
meridional circulation in the radiative zones is governed by a system of
five first order differential equations. These equations are obtained by
splitting Eq.~(\ref{AM}) into first order equations that are complemented
by Eq.~(\ref{ThermalWind}) and Eq.~(\ref{Lambda}). We use a Newton-Raphson
relaxation method \citep{Heyney64} to solve these equations. In
convection zones we assume solid-body rotation ($\Omega(r) = cst$) and do
not solve the aforementioned system. The boundary conditions at the
radiative zone limits are the same as those described in \citet{PA03}.
  
We choose initial velocities on the ZAMS that reflect an average velocity
for main sequence B and F stars \citep{F82,Abt02,G93}. 
The adopted initial velocity on the ZAMS are $\upsilon_{\rm ZAMS} =
300$~\kms{} and $\upsilon_{\rm ZAMS} = 100$~\kms{}
for the 20 M$_\odot$ and the 1.5 M$_\odot$ respectively.

The evolution of the surface velocity can be affected by the magnetic
torques exerted at the surface of the star during its early evolution
\citep{Schatzman62}. This magnetic braking is generally associated with the
presence of a convective envelope ; it will thus only be applied to the
low-mass star model, following \citeauthor{K88}'s \citeyearpar{K88}
  formalism as described in \citet{PA03}. In the case of massive stars,
it is only very recently that magnetic fields have been detected in normal
O and B stars, and they appear to be very weak or nonexistent
\citep{Bouret08,Schnerr08}.  Moreover, O and B stars with measured
$\upsilon \sin i$ appear to be fast rotators. For these reasons, we have
decided not to account for magnetic braking in the 20~\Ms{} model.

In the following sections, we present new diagnosis tools to better
  understand and quantify the processes
responsible for the building up of differential rotation in stellar interiors.

\subsection{Rotational profiles and differential rotation}\label{sec:omdiff}

\subsubsection{20 M$_\odot$ model}

\begin{figure}[tbp]
\begin{center}
\includegraphics[width=0.35\textwidth]{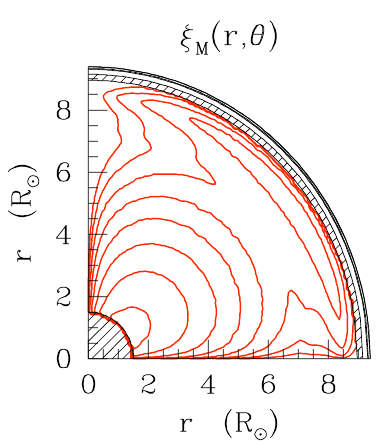}
\caption{Meridional circulation currents in the 20~\Ms{} model when
  $X_\text{c} = 0.32$.  Solid thin (red) lines indicate counterclockwise
  circulation ($U_2<0$), which carries angular momentum outwards. Hatched
  regions correspond to convection zones.}
\label{fig:U2xsi20}
\end{center}
\end{figure}

\begin{figure}[tbp]
\begin{center}
\includegraphics[width=0.35\textwidth]{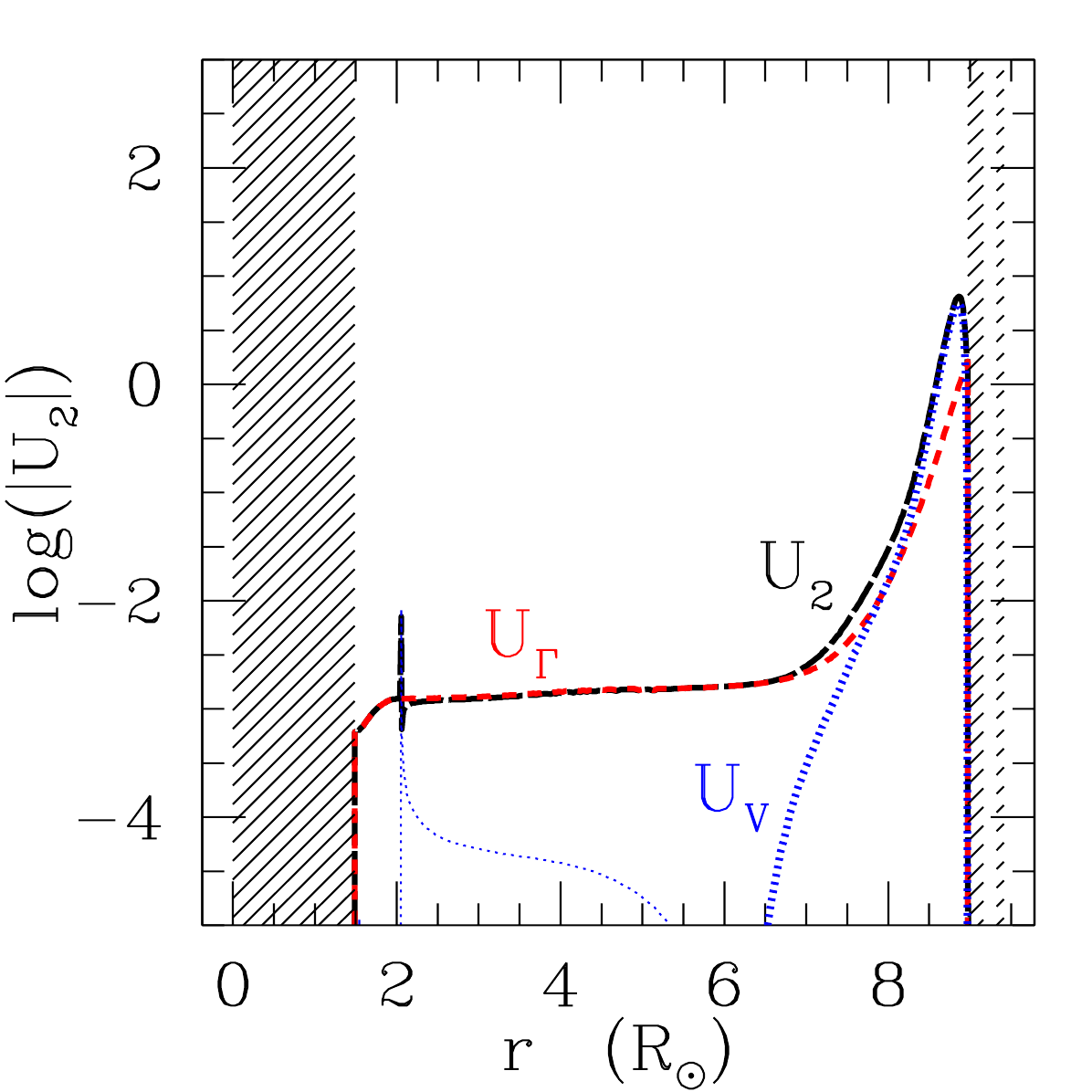}
\caption{Decomposition of the meridional circulation based on the equation
  of angular momentum evolution in the 20~\Ms{} star, when $X_\text{c} =
  0.32$. Hatched regions correspond to convection zones.  $U_\Gamma$
  (short-dashed red line, see Eq.~\ref{eq:ucirc}), $U_2$ (meridional
  circulation, in long-dashed black line), and $U_V$ (turbulent
  transport, in dotted blue line) are presented on a logarithmic
  scale. The thin part of the curves means that the plotted quantity is
  positive and the bold parts that it is negative.}
\label{fig:U2recomp20}
\end{center}
\end{figure}

The 20~\Ms{} model does not undergo magnetic braking and thus
remains a fast rotator during its main sequence (MS) evolution, with
$\upsilon_{\rm surf}\simeq 250$~\kms.  Similarly to what was obtained by
\citet{MM00}, differential rotation rapidly develops in the
interior, as firstly meridional flow transports
angular momentum inwards. 
After a transient phase that lasts $\sim 1\,$~Myr following the arrival on
the ZAMS, the morphology of the rotation profile is not significantly
affected until the star reaches the turn-off, as shown in
Fig.~\ref{fig:omega} (left panel). 

Above the radial coordinate 5~\Rs{} (resp. 6~\Rs{} and 10~\Rs{}) for
$X_\text{c} = 0.675$ (resp. $X_\text{c} =0.32$ and $X_\text{c}=0.0$) the
angular velocity gradient becomes positive due to a very efficient
transport of angular momentum by the meridional circulation in a region
where the density is low (see \S~\ref{sec:circ}).  This inversion of
the angular velocity gradient should be counteracted by the shear
turbulence that tends to smooth out this profile. However, the shear can only
operate if the flow is turbulent, i.e. if the Reynolds condition $\nu_V \ge
\nu Re_c$ is satisfied where $Re_c \simeq 1$ is the critical Reynolds
number and $\nu_V$ and $\nu$ the shear and microscopic viscosities,
respectively.  Below the convective envelope (where
$\partial_{r}{\overline\Omega} \simeq 0$), this condition is not fulfilled,
and the shear is unable to connect the region with positive angular
velocity gradient with the one where $\partial_{r}{\overline\Omega}=0$ (see
the dip in the profile of $D_\nu$ in Fig.~\ref{fig:diff20}). This explains
the persistence of this positive gradient. Computations done without taking
into account the Reynolds criterion show that this positive gradient is
reduced (by a factor 3 to 5) but is still present, although at a lower
level. It should be noted that this inversion has already been obtained by
\citet{TZMM97} in the asymptotic regime for a 9~\Ms{} with an
equatorial angular velocity of 425~\kms{}. This feature can be linked to
the Gratton-\"Opik term in the meridional circulation (see \S~3.4.1).

The decrease in the angular velocity during the MS evolution is due to two
factors: (1) mass and angular momentum losses by the stellar winds and (2)
structural changes that lead to an overall expansion of the star. It should
be noted however that up to the point where $X_\text{c} \simeq 0.1$, the
contrast in $\Omega$ between the center and the surface is small, not
exceeding 20\% .

\subsubsection{1.5 M$_{\odot}$ model}

Contrary to the 20~\Ms{} star, the 1.5~\Ms{} star undergoes magnetic
braking at its arrival on the ZAMS, resulting in a strong extraction
of angular momentum, as shown in \citet{TC98} and
\citet{PA03,PCTS06}. During the first 800~Myr the surface equatorial
velocity drops from 100~\kms{} to 78~\kms{}, in agreement with observations
in open clusters like the Hyades \citep[i.e.,][]{G93}. Beyond this point,
the braking becomes less efficient and the surface rotational velocity
continues to decrease slowly until the star reaches the turn-off with
$\upsilon_{\rm surf} \simeq 53$~\kms{} at age 2.35~Gyrs.

The right panel of Fig.~\ref{fig:omega} presents the evolution of the
angular velocity profile on the MS. When $X_\text{c} = 0.675$, i.e. when
the star is $\sim 200$~Myr old, the core already rotates 1.5 times faster
than the surface. While the angular velocity in the convective core slowly
increases, the extraction of angular momentum at the surface slows down the
surface layers and differential rotation increases during the MS. When the
convective core disappears at central H-exhaustion, the contraction of the
inner shells produces a much steeper $\Omega$-gradient (dashed profile in
right panel, Fig.~\ref{fig:omega}) with $\Omega_{\rm center} = 14
\times \Omega_{\rm surf}$.

\subsection{Meridional circulation}\label{sec:circ}
In this section, we isolate the processes which are at the origin
    of meridional circulation.

\subsubsection{20 M$_{\odot}$ model}

Figure~\ref{fig:U2xsi20} shows the meridional flows in the radiative
interior of the 20~\Ms{} when the central hydrogen mass fraction reaches
$X_\text{c} = 0.32$. They are represented in terms of the stream function
$\xi_{\rm M}\left(r,\theta\right)$ (see Eq.~\ref{eqUloop}). The circulation
consists of a single counterclockwise loop, by which matter is transported
inward along the rotational axis, and is conveyed outward in the equatorial
plane. These flows thus extract angular momentum from the deep interior to
the surface. Matter is conserved in the meridional circulation currents and
$\overline\rho\, ||\vec{\mathcal U}_{\rm{M}}||=\text{constant}$ along the
flux lines (shown in Fig.~\ref{fig:U2xsi20}). The abrupt jump in density
occurring near the surface produces an increase in the meridional
circulation velocity as illustrated by Fig.~\ref{fig:U2recomp20}. This in
turn forces the inversion of the angular velocity gradient mentioned in
\S~\ref{sec:omdiff} via the Gratton-\"Opik term ($\propto -
{\overline\Omega^2}/{2\pi G \overline\rho}$, see Eq.~\ref{TB} in Appendix
~\ref{ap:A2}). The velocities are maximum in the low density regions
(i.e. below the surface) where they reach $||{\vec{\mathcal U}}_{\rm
  M}(r,\theta)||$ $\approx$ 1.6~m~s$^{-1}$.

\begin{figure*}[tbp]
\centering
\includegraphics[width=0.925\textwidth]{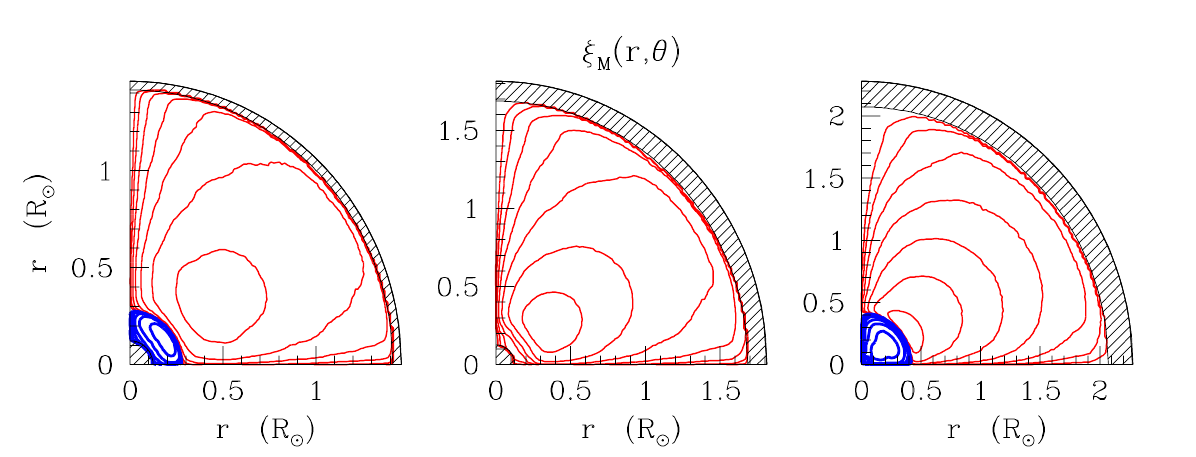}
\caption{Meridional circulation flow in the 1.5~\Ms{} model when
  $X_\text{c} = 0.675$ (\emph{left panel}), 0.32 (\emph{middle panel}) and
  0.0 (\emph{right panel}). Solid thin (red) lines indicate counterclockwise
  circulation ($U_2 < 0$) and solid bold (blue) lines clockwise circulation
  ($U_2 > 0$). In this model, the outer cell is turning counterclockwise
  allowing the extraction of angular momentum by the wind. Hatched regions
  indicate convection zones.}
\label{fig:2D1.5xsi}
\end{figure*}

\begin{figure*}[tbp]
\begin{center}
\includegraphics[width=0.925\textwidth]{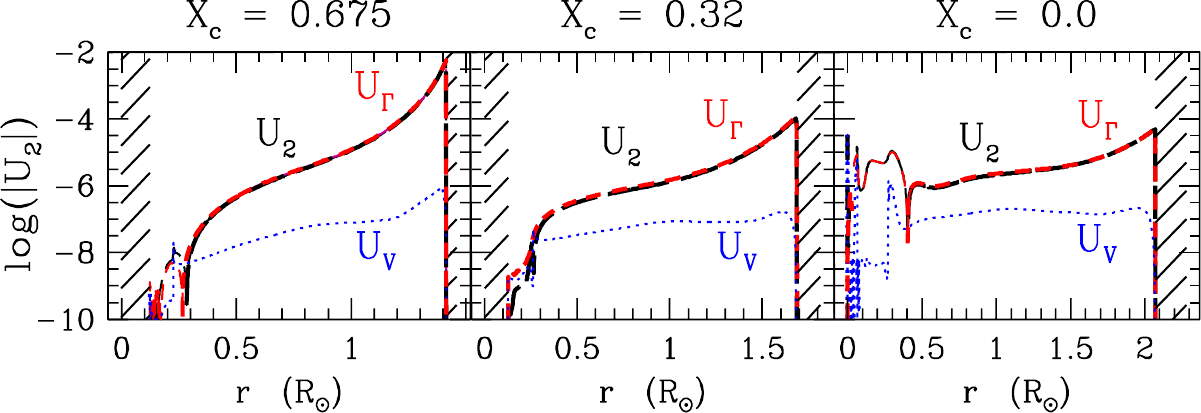}
\caption{Decomposition of the vertical component of the meridional
  circulation $U_2$ according to Eq.~(\ref{eq:ucirc}) for the 1.5~\Ms{}
  model at the three different epochs indicated on top of each panel. The
  term due to angular momentum transport, ${\rm U_\Gamma}$, is shown in
  dashed red line, the one due to vertical shear, ${\rm U_V}$, is in dotted
  blue line. The value of ${\rm U_2}$ obtained in the simulation is given
  in long-dashed black line. Thin and thick lines represent respectively
  positive and negative values.}
\label{fig:U2recomp1.5}
\end{center}
\end{figure*}

After a short adjustment phase, which depends on the initial conditions
that have been imposed on the ZAMS (uniform rotation), the meridional
circulation settles into a counterclockwise regime, as was noticed by
\citet{MM00}. This flow thus carries angular momentum
outwards, in order to compensate the structural changes, i.e. the
inflation of the envelope and the contraction of the core.

Figure~\ref{fig:U2recomp20} shows the {\em a posteriori} reconstruction of
the meridional circulation based on the integration of the equation for the
evolution of angular momentum (Eq.~\ref{eq:ucirc}), at
$X_\text{c}=0.32$. $U_\Gamma$ represents the angular momentum extraction
and $U_\text{V}$ that due to the shear. In the inner part of the radiative
zone, below 7~\Rs, the meridional circulation is powered by the extraction
of angular momentum, as $|U_\Gamma| \gg |U_\text{V}|$. Near the surface,
$U_\text{V}$ becomes negative and of the same order as $U_\Gamma$. This
inversion of $U_\text{V}$ is correlated with the sign change of the angular
velocity gradient since $U_\text{V}=-{5 \nu_v}\partial_{r}\ln {\overline\Omega}$.
$U_\Gamma$ may be considered as a measure of the departure from a
stationary regime, where angular momentum transport by shear ($U_\text{V}$)
and meridional circulation ($U_2$) compensate each other. This balance is
barely achieved in the uppermost part of the star (above 8~\Rs{}), but the
bulk of the radiative interior is not in a stationary state, as already
pointed out in \citet{TZMM97} and \citet{MM00} on the basis
of an asymptotic analysis. Here, we confirm this result but also track down
the importance of the shear versus angular momentum transport in shaping
the meridional circulation.

\subsubsection{1.5 M$_{\odot}$ model}

The wind-driven meridional flows are depicted in Fig.~\ref{fig:2D1.5xsi}
for the low-mass star, at the three evolutionary stages indicated in
Fig.~\ref{fig:HR}. Compared to the 20~\Ms{} model, the topology of the
meridional circulation is more complex. In panels 1 and 3, the circulation
presents two loops: a small clockwise loop in the central regions and a
large counterclockwise loop connecting the interior to the surface. As in
the massive star model, the counterclockwise loop carries angular momentum
outward and is driven by the strong extraction of angular momentum (here
largely due to the magnetic braking). The clockwise loop corresponds to
matter flowing from the equator to the pole, which results in deposition of
angular momentum in this region. Such loops appear only in the center and
do not persist over the whole evolution on the main sequence.

At the beginning of the MS evolution, the meridional circulation velocity
is maximal in a thin layer below the surface as a result of the applied
magnetic torque. The amplitude of the meridional circulation velocity
$||U_2||$ does not exceed 0.5~m~s$^{-1}$ at its maximum and is of the order
of $10^{-4}$ in the bulk of the radiative zone. As the star spins down, the
braking becomes less efficient and the meridional circulation weakens
by 3 orders of magnitude. The large difference in meridional velocity between
the low-mass and the massive star is due to the $L/M$ factor (Eq.~\ref{eq:Mer}), which increases with
stellar mass, and to the higher initial angular velocity of the 20~\Ms{}
star.

Figure~\ref{fig:U2recomp1.5} is similar to Fig.~\ref{fig:U2recomp20} but
  for the 3 models selected during the MS evolution of the 1.5~\Ms{} star. As
for the 20~\Ms{} model, the meridional circulation is driven by the local
loss of angular momentum ($U_\Gamma$ term which matches the $U_2$
profile), the contribution of the vertical shear ($U_{V}$) being always
much smaller, except in the central regions where it can be of the same order
as $U_\Gamma$. This figure confirms that the inward transport of
angular momentum (clockwise meridional circulation loops) is associated
with a  positive value of $U_\Gamma$, corresponding to a local gain of angular momentum (see
Eq.~\ref{eq:ucirc}). In most of the radiative zone, $U_\text{V} \ll U_2$,
and contrarily to what was found in the massive star model, the circulation in
the upper part of the radiative zone never approaches the stationary
regime. As shown by \citet{TZMM97}, the extraction of angular momentum
by magnetic braking together with the very small meridional velocities prevent the inversion of
the angular velocity gradient in the regions of low density as was the case
in the massive star model. Consequently, $U_\text{V} \propto
- \partial_{r}\overline\Omega$ remains always positive. Besides, during
the main sequence the differential rotation remains too small to ensure an
efficient transport of angular momentum by the shear turbulence.

\begin{figure}[tbp]
\begin{center}
\includegraphics[width=0.325\textwidth]{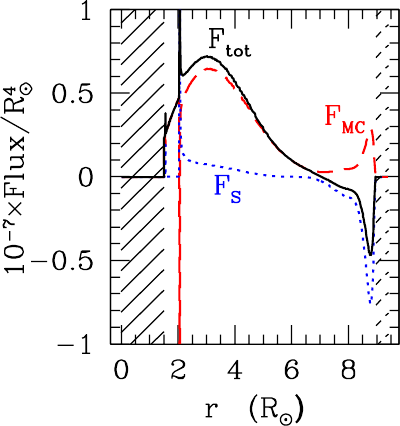}
\caption{The total flux of angular momentum (solid black line) decomposed
  into its meridional circulation $F_{\rm
    MC}\left(r\right)=-\frac{1}{5}{\overline\rho}
  r^4\overline{\Omega}U_{2}$, (dashed red line) and turbulence $F_{\rm
    S}\left(r\right)=-{\overline\rho}\nu_{v}r^4\partial_{r}\overline{\Omega}$
  components (dotted blue line) for the 20~\Ms{} model when $X_\text{c} =
  0.32$.  The fluxes are rescaled by~${R_{\odot}^{4}}$.}
\label{fig:fluxAM20}
\end{center}
\end{figure}

\subsection{Angular momentum transport}

We already compared in \S~\ref{sec:circ} the respective contributions of
meridional circulation and turbulence in the transport of angular
momentum, in agreement with results from the literature.
We shall now examine the problem from a different perspective,
based on the fluxes carried by the two processes (Eq.~\ref{fluxAM}).

\subsubsection{20 M$_{\odot}$ model}

Figure~\ref{fig:fluxAM20} shows the two components of the angular momentum
flux associated with the meridional circulation and shear. 
In this model the
meridional circulation presents only one counterclockwise loop so the
transport occurs in the same direction in the whole radiative zone, and
carries angular momentum from the core to the surface ($F_{\rm MC} >
0$). Below $r = 6\,R_\odot$ the turbulent shear works in the same
direction ($F_{\rm S} > 0$), but its contribution is
negligible compared to that of the meridional circulation. Above $6\,
R_\odot$, the angular velocity gradient changes sign and shear takes over
the advection of angular momentum.

$F_{\rm S}$ and $F_{\rm MC}$ are of opposite sign below the convective
envelope during the MS evolution, and grow in amplitude due to the radial
expansion of the outer low-density layers (see Eqs.~\ref{eq:Fmc} and
\ref{eq:FS}). This configuration tends to impose a stationary
circulation in this region. On the other hand, these fluxes have the
same sign deeper in the interior, and act jointly to compensate for 
the structural changes of the star by redistributing the total angular
momentum.

\begin{figure*}[tbp]
\begin{center}
\includegraphics[width=0.925\textwidth]{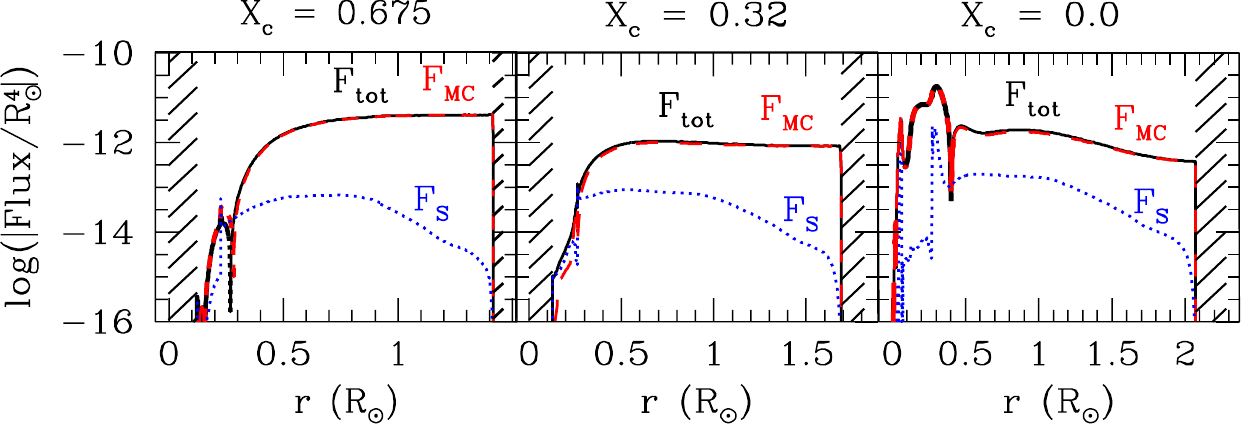}
\caption{Same as Fig.~\ref{fig:fluxAM20} for the 1.5~\Ms{} model at the
  three selected evolutionary stages indicated by $X_\text{c}$. In this
  graph, the logarithm of the absolute value of the flux is plotted and
  rescaled by~$R_{\odot}^{4}$. Thin and bold lines represent positive and
  negative fluxes respectively.}
\label{fig:fluxAM1.5}
\end{center}
\end{figure*}

\subsubsection{1.5 M$_{\odot}$ model}

In the low-mass model, the meridional circulation dominates at all times  the angular
momentum transport and $F_{\rm MC} \simeq F_{\rm Tot}$ on the
main sequence (Fig.~\ref{fig:fluxAM1.5}). The bold parts of the curves
indicate positive, outward fluxes that are associated with the
counterclockwise meridional currents described in Fig.~\ref{fig:2D1.5xsi}.

Unlike in the massive star model, shear extracts angular momentum
everywhere because of the monotonous decrease of $\Omega$ with
radius. Nevertheless, its contribution remains negligible. At the turn-off,
the meridional circulation and angular velocities have dropped in the outer
parts of the radiative zone because of the braking of the surface layers
and structural changes. This weakens $F_{\rm MC}$ while $F_{\rm S}$ remains
almost unaltered. In the central regions, both $\Omega$ and $\partial_r
\Omega$ substantially increase leading to a concomitant rise in $F_{\rm
  MC}$ and $F_{\rm S}$. The increase and sign change of $F_{\rm MC}$
between $r = 0.05$~\Rs{} and $r = 0.4$~\Rs{} is partly due to the
off-center displacement of the nuclear energy production region when H is
exhausted in the core.

In summary, our diagnostic
  tools applied on two well-known cases
allow us to give quantitative estimates on the main properties
    of angular momentum transport  already discussed
    in the literature \citep[see e.g.,][]{MM00,PA03}. In low-mass stars,
  angular momentum transport is ensured by meridional currents that are
primarily generated by the action of the applied torques resulting from the
action of magnetic braking. In the massive star, owing to the differences
in structure (density stratification, radius) and in surface velocities,
the gradient of angular velocity becomes positive in the outer layers of
the star and there shear turbulence takes over the transports of angular
momentum.

\subsection{Thermal relaxation}

Let us now analyze how thermal relaxation is achieved, as described by the
heat equation (\ref{eq:Mer}), which rules the temperature perturbations.
As explained previously, the meridional circulation is mainly driven
 by the extraction of angular momentum due to angular momentum losses
at the surface (i.e. by the external torques) and to structural
readjustments (such as the expansion of the star on the MS). This
circulation advects specific entropy, which perturbs the thermal
equilibrium and induces temperature fluctuations described by
$\Psi_{2}$. These are linked to the differential rotation through the
baroclinic equation (\ref{ThermalWind}), so that the heat equation is
actually tightly coupled with the equation of angular momentum transport
(\ref{AM}).
We examine how this works in our models.

\subsubsection{20 M$_{\odot}$ model}

\begin{figure*}[tbp]
\begin{center}
\includegraphics[width=\textwidth]{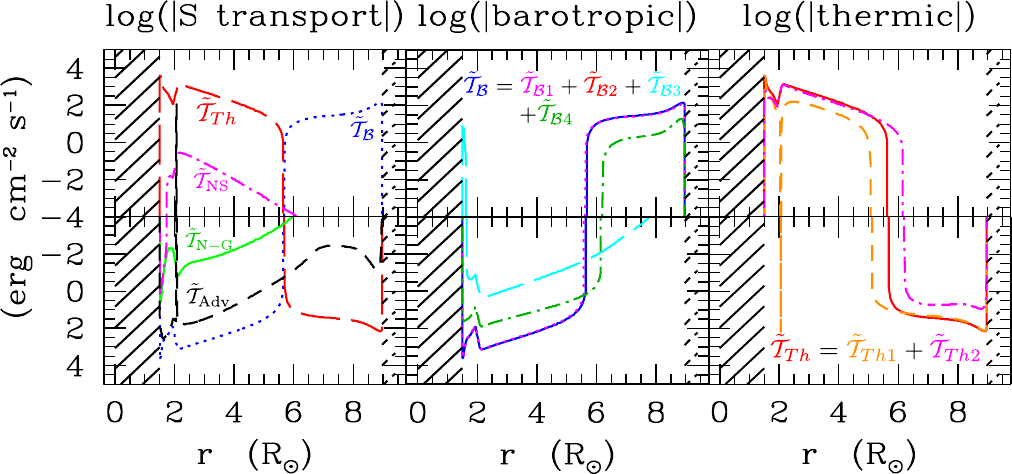}
\caption{Representation of the different components of the heat equation
  (Eq.~\ref{eq:Mer}) for the 20~\Ms{} model when $X_\text{c} = 0.32$
  (left panel).  The upper panel shows log$|$S$|$ for $S > 0$ while
  the lower panel shows log$|$S$|$ for $S < 0$. Shown are the
  non-stationary (NS, dot long-dashed magenta line), barotropic
  (${\mathcal B}$, dotted blue line), baroclinic ($Th$ long-dashed red
  line), advective (dashed black line) terms and the contribution due to
  the nuclear and gravitational energy variations (N-G, solid green
  lines). The central panel shows the components of the barotropic
    term:  $\tilde{\mathcal T}_{\mathcal B}$ (solid line), $\tilde{\mathcal
      T}_{\mathcal B1}$ (dotted line),
    $\tilde{\mathcal T}_{\mathcal B3}$ (long-dashed line), $\tilde{\mathcal
      T}_{\mathcal B4}$ (dotted-dashed line). The right panel shows the
    thermic components: $\tilde{\mathcal T}_{Th}$ (solid), $\tilde{\mathcal T}_{Th1}$ (dashed
    line) and $\tilde{\mathcal T}_{Th2}$ (dotted line). }
\label{fig:heat20}
\end{center}
\end{figure*}

\begin{figure}[h!]
\begin{center}
\includegraphics[width=0.32\textwidth]{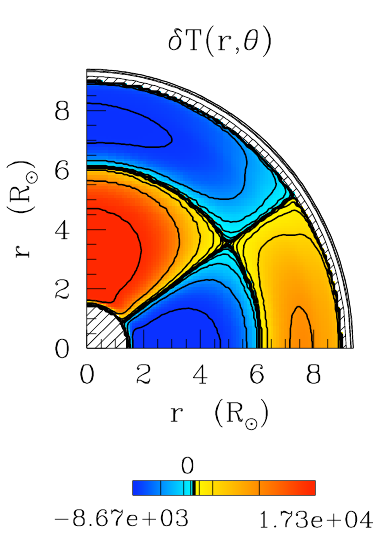}
\caption{Two-dimensional  reconstruction  of  the temperature  fluctuations
  $\delta T  = \overline T \Psi_2  P_2(\cos \theta)$ in  the 20~\Ms{} model
  when  $X_\text{c}  = 0.32$.  The  light  (blue)  parts are  for  negative
  fluctuations and the darker (red) are for positive ones.}
\label{fig:deltaT20}
\end{center}
\end{figure}

\begin{figure*}[ht!]
\begin{center}
\includegraphics[width=0.925\textwidth]{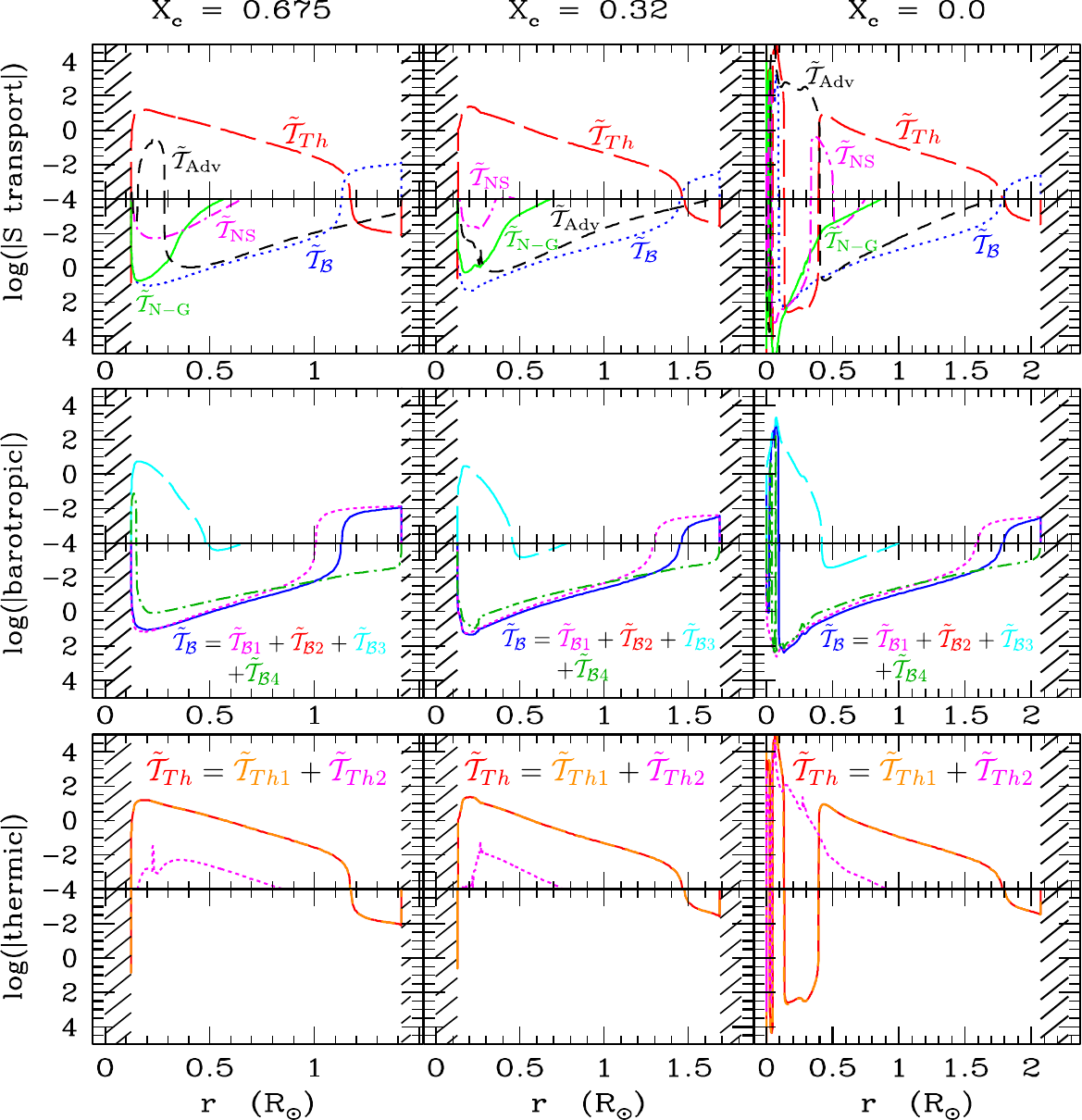}
\caption{Same as Fig.~\ref{fig:heat20} for the 1.5~\Ms{} model at the three
  evolutionary points indicated by $X_\text{c}$. All quantities are
    shown in erg~cm$^{-2}$~s$^{-1}$.}
\label{fig:heat1p5}
\end{center}
\end{figure*}

\begin{figure*}[ht!]
\centering
\includegraphics[width=0.925\textwidth]{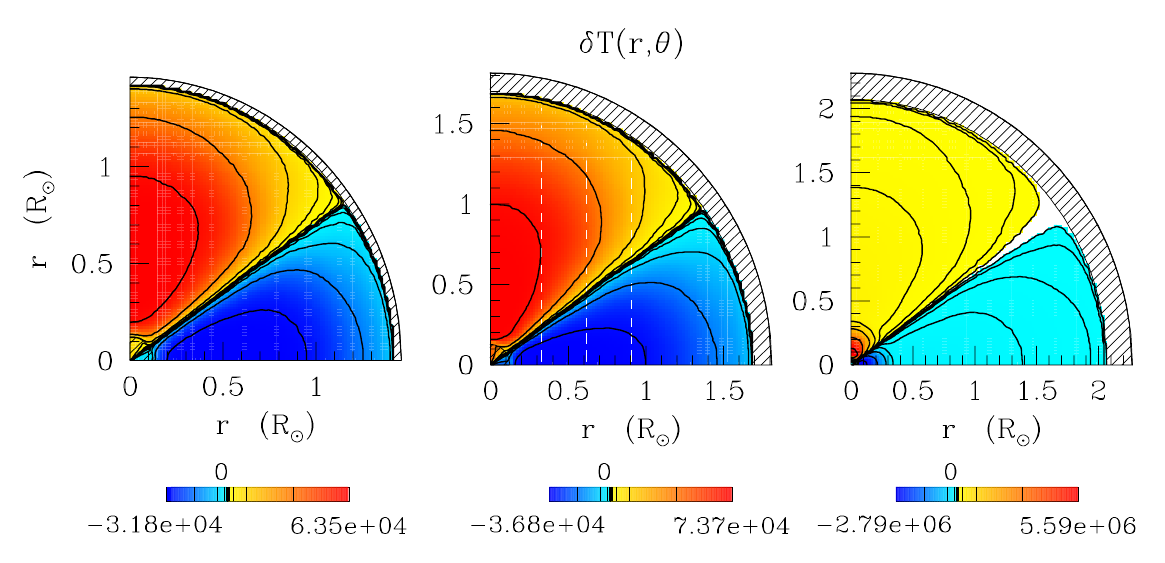}
\caption{The temperature fluctuation $\delta T = \overline T \Psi_2 P_2(cos
  \theta)$ in the 1.5~\Ms{} model at three different points on the main
  sequence, when the hydrogen mass fraction in the center $X_\text{c}$ =
  0.675 ({\em left panel}), $X_\text{c}$ = 0.32 ({\em middle panel}) and
  $X_\text{c}$ = 0 ({\em right panel}).}
\label{fig:2D1.5T}
\end{figure*}
Each term of Eq. (\ref{RelaxThEq}) is
represented in Fig. \ref{fig:heat20} (left panel). The plot is
separated in two panels, the upper and lower ones showing the positive
and negative parts of each profile respectively, on a logarithmic
scale. The differential rotation adjusts itself so that the advection of
entropy by the meridional circulation ($\tilde{\mathcal T}_{\rm Adv}$) is almost exactly compensated by
the thermal readjustment via the barotropic ($\tilde{\mathcal T}_{\mathcal B}$) and thermic
($\tilde{\mathcal T}_{\rm Th}$) terms. The other terms, associated
  with the non-stationarity ($\tilde{\mathcal T}_{\rm NS}$) and with the
  nuclear and gravitational energy generation ($\tilde{\mathcal T}_{\rm N-G}$), are
  several orders of magnitude smaller and thus can be neglected during the
  main sequence evolution. From this simplification, splitting the divergence
  of the radiative flux (third term in the right-hand side of
  Eq.~\ref{eq:Mer}) into barotropic and thermic (baroclinic) components thus
  appears to be somewhat artificial.

The temperature perturbations $\Psi_2$ quickly relax to
the asymptotic regime, which involves only the source ($\tilde{\mathcal T}_{\rm Adv}$) and diffusion
($\tilde{\mathcal T}_{\mathcal B}+\tilde{\mathcal T}_{\rm Th}$) terms
(see also Appendix B for description of these terms), and
  Eq.~(\ref{eq:Mer}) gives: 
\begin{equation} 
  \vec\nabla\cdot\left(\chi\vec\nabla T\right)-\vec\nabla\cdot\vec F_{H}
  \approx  -{\overline\rho}{\overline T}U_{r}\partial_{r}{\overline S}
\end{equation}
and
\begin{equation} 
\tilde{\mathcal T}_{2,{\mathcal
        B}}+\tilde{\mathcal T}_{2,\rm Th} \approx{\overline \rho}{\overline 
    T}C_{p}\frac{N_{T}^{2}}{{\overline g}\delta}U_{2}.
\label{eq:Tb}
\end{equation}

To explicitly disentangle the role played by each component in shaping
  the entropy transport, we split the barotropic term as follow: 
   \begin{equation}
   \label{eq:splitB}
   \tilde{\mathcal T}_{{\mathcal B}} = \tilde{\mathcal T}_{{\mathcal B1}}+\tilde{\mathcal T}_{{\mathcal B2}}+\tilde{\mathcal T}_{{\mathcal B3}}+\tilde{\mathcal T}_{{\mathcal B4}},
   \end{equation}
where
\begin{equation}
  \label{eq:splitB1}
  \tilde{\mathcal T}_{{\mathcal B1}} = {\overline \rho}\frac{L}{M} \frac23
         \left(1-\frac{\overline\Omega^2}{2\pi
             G\overline{\rho}}\right)\overline\Omega^{2}\partial_{r}
         \left(\frac{r^2}{\overline g}\right),
\end{equation}
\begin{equation}
  \label{eq:splitB2}
  \tilde{\mathcal T}_{{\mathcal B2}} = - {\overline
         \rho}\frac{L}{M} \frac49
         \frac{\rho_m}{\overline\rho}\left(\varphi\Lambda_{2}-\delta\Psi_{2}\right)\overline\Omega^{2}\partial_{r} \left(\frac{r^2}{\overline
g}\right),
\end{equation}
\begin{equation}
  \label{eq:splitB3}
  \tilde{\mathcal T}_{{\mathcal B3}} = -{\overline
         \rho}\frac{L}{M} \frac23 \frac{\left(\overline{\epsilon}+\overline{\epsilon}_{\rm
               grav}\right)}{\epsilon_{m}}
      \overline\Omega^{2}\partial_{r} \left(\frac{r^2}{\overline
g}\right),
\end{equation}
and 
\begin{equation}
  \label{eq:splitB4}
  \tilde{\mathcal T}_{{\mathcal B4}} = -{\overline
         \rho}\frac{L}{M}\frac23\frac{\rho_m}{\overline\rho}\left(\varphi\Lambda_{2}-\delta\Psi_{2}\right).
\end{equation}
$\tilde{\mathcal T}_{{\mathcal B1}}$ represents the "pure" barotropic term
(which does not depend on $\partial_r{\overline\Omega}$), while 
$\tilde{\mathcal T}_{{\mathcal B2}}$ and $\tilde{\mathcal T}_{{\mathcal
    B4}}$ are the first and second baroclinic corrections due to the 
differential rotation (cf. Eq. \ref{ThermalWind}). Finally,
$\tilde{\mathcal T}_{{\mathcal B3}}$ is associated with the energy
production on the isobar.

Next, the thermic term is decomposed as follows:
\begin{equation}
    \label{eq:splitTh}
    \tilde{\mathcal T}_{\rm Th} = \tilde{\mathcal T}_{{\rm Th}1}+\tilde{\mathcal T}_{{\rm Th}2},  
\end{equation}
where
\begin{equation}
  \label{eq:splitTh1}
  \tilde{\mathcal T}_{{\rm Th}1} ={\overline
         \rho}\frac LM 
       \left(
         \frac{\rho_{m}}{\overline{\rho}}\frac{r}{3}\partial_{r}{\cal{A}}_{2}\left(r\right)
         -\frac{2 H_{T}}{3 r}{\Psi_{2}}
       \right)
,
\end{equation}
and 
\begin{equation}
  \label{eq:splitTh2}  
  \tilde{\mathcal T}_{{\rm Th}2} =-{\overline
         \rho}\frac{L}{M}\frac{2 H_{T}}{3 r}
       \frac{D_{h}}{K_{T}} \Psi_{2}
.
\end{equation}
The term $\tilde{\mathcal T}_{{\rm Th}1}$ comes from the diffusive laplacian acting on $\Psi_2$ while
$\tilde{\mathcal T}_{{\rm Th}2}$ is related to the heat flux carried by the
horizontal turbulence ($\vec\nabla\cdot\vec F_{H}$).

Middle and left panels of Fig.~\ref{fig:heat20} display these components
for the barotropic ($\tilde{\mathcal T}_{{\mathcal B}}$) and thermic
($\tilde{\mathcal T}_{\rm Th}$) terms. We first note that the behaviour of
$\tilde{\mathcal T}_{{\mathcal B}}$ closely follows that of
$\tilde{\mathcal T}_{{\mathcal B1}}$. This result is expected as the terms
$\tilde{\mathcal T}_{{\mathcal B2}}$ and $\tilde{\mathcal T}_{{\mathcal
    B4}}$ are driven by the T- and $\mu$-fluctuations, which remain small
in main sequence stars (see below). Moreover the term $\tilde{\mathcal
  T}_{{\mathcal B3}}$ scales with
$(\overline{\epsilon}+\overline{\epsilon}_{\rm grav})/\epsilon_{m}$ (i.e.,
the ratio of energy production at a given radius with respect to the total
energy enclosed in the sphere of this radius), which remains small as the
(nuclear) energy production is mainly located in the convective core. It
should also be noted that the relative strength of the two baroclinic
corrections scales with $\overline\Omega^{2}\partial_{r}
\left(\frac{r^2}{\overline g}\right)$, which is always very small (between
$10^{-14}$ and $10^{-10}$ in the whole radiative region and mainly varies
via the ${\overline\Omega}^2$ dependence).  This explains why the term
$\tilde{\mathcal T}_{{\mathcal B2}}$ does not appear in
Fig.~\ref{fig:heat20} (middle panel) and can be neglected during the main
sequence evolution.

The thermic term ($\tilde{\mathcal T}_{\rm Th}$) requires more attention as
no specific component dominates in the whole radiative layer. Below the
surface and above the convective core the term $\tilde{\mathcal T}_{{\rm
    Th}1}$ dominates while $\tilde{\mathcal T}_{\rm Th}$ is driven by
$\tilde{\mathcal T}_{{\rm Th}2}$ in the center of the radiative zone. As it
will be discussed in the next section, this result strongly depends on the
$D_h/K_T$ ratio.

Figure~\ref{fig:deltaT20} presents the two-dimensional reconstruction of
the temperature perturbations in the meridional plane, i.e.  $\delta
T\left(r,\theta\right)= \Psi_2(r){\overline T}(r) P_2(\cos \theta)$. These
are directly connected to the behaviour of $\overline\Omega$-gradient: the
temperature is higher at the poles (closer to the rotational axis) than at
the equator where $\partial_{r}{\overline\Omega} < 0$, i.e. where the
angular velocity decreases with increasing radius. Near the surface, the
situation is reversed since $\partial_{r}{\overline\Omega}$ is positive
there.

\subsubsection{1.5 M$_{\odot}$ model}

Figure~\ref{fig:heat1p5} is similar to Fig.~\ref{fig:heat20} for the
1.5~\Ms{} model, but contrarily to the 20~\Ms{} model, $\tilde{\mathcal
  T}_{\rm Adv}$ is now of the same order of magnitude as $\tilde{\mathcal
  T}_{\mathcal B}$ and $\tilde{\mathcal T}_{\rm Th}$. This is because
magnetic braking forces a larger extraction of angular momentum. Let us
note that the reversal of the meridional circulation shown in
Figs.~\ref{fig:2D1.5xsi} and~\ref{fig:U2recomp1.5} is evidenced by the sign
change of $\tilde{\mathcal T}_{\rm Adv}$. Also in contrast with the massive
star, the nuclear and gravitational term, $\tilde{\mathcal T}_{\rm N-G}$,
becomes comparable to the advective one near the convective core, and
should thus not be neglected. Finally, except near turn-off where it
becomes important in the interior, the non-stationary term,
$\tilde{\mathcal T}_{\rm NS}$, is generally several orders of magnitudes
smaller than the other terms during most of the main sequence.

Equation~(\ref{RelaxThEq}) is thus left with four terms balancing each
other. The thermic (baroclinic) and barotropic terms have similar
amplitudes and opposite signs and their sum balances the sum of the
advective term and the nuclear and gravitational heating:
    \begin{equation}
  \tilde{\mathcal T}_{\mathcal B} + \tilde{\mathcal T}_{\rm Th} \simeq
  \tilde{\mathcal T}_{\rm adv}+\tilde{\mathcal T}_{\rm N-G}.
\end{equation}
Let us note finally that this situation is not valid anymore at the
turn-off as seen in the third column of Fig.~\ref{fig:heat1p5}. There, the
non-stationary terms become dominant in the deep interior and must not be
neglected anymore for the following evolutionary phase.

Figure~\ref{fig:heat1p5} also displays the components of the barotropic and
thermic term in the same way as in Fig.~\ref{fig:heat20}. In most of the
radiative interior $\tilde{\mathcal T}_{{\mathcal B}}$ follows the
barotropic component, $\tilde{\mathcal T}_{{\mathcal B1}}$, except below
the surface where the contribution of the second baroclinic correction
$\tilde{\mathcal T}_{{\mathcal B4}}$ becomes relevant. As in the 20~\Ms{}
model, $\tilde{\mathcal T}_{{\mathcal B3}}$ is negligible during the main
sequence. However in the last model ($X_c = 0.0$), this term dominates in
the center due to the release of (a) nuclear energy by the hydrogen-burning
shell and (b) gravitational energy by the contracting core.

Contrary to the 20~\Ms{}, the heat diffusion, $\tilde{\mathcal T}_{{\rm
    Th}2}$ ($\propto D_h/K_T$), plays no role in the thermic term,
$\tilde{\mathcal T}_{\rm Th}$, which is solely driven by the laplacian
component, $\tilde{\mathcal T}_{{\rm Th}1}$. The 1.5~\Ms{} model has been
computed with the \citet{zahn92} prescription for the horizontal turbulent
diffusion coefficient $D_h$ leading to $D_h/K_T \simeq 10^{-4} - 10^{-3}$,
whereas this ratio amounts to about 10 for the 20~\Ms{}, where the
\citet{MPZ04} prescription was used (see \S~3.5).  At the center of the
last shown model, the laplacian component presents a very complex behaviour
as it is sensitive to the first, second and third radial derivatives of
$\overline\Omega$. Fig.~\ref{fig:omega} reveals that the profile of
$\overline\Omega$ presents some rapid variations in this region which
explain this intricate feature.

Figure~\ref{fig:2D1.5T} displays the temperature fluctuations which
essentially follow the gradient of $\overline\Omega$
as in the 20~M$_{\odot}$ case. The main difference
is due to the presence of a second circulation loop. The amplitude of
  the fluctuations remains globally of the same while the star is on the
MS, and amount to about a few percents. 

\subsubsection{Possible simplifications}

From the previous analysis of the 1.5 and 20~\Ms{} models, during the
quiescent main sequence evolution, the relaxation of $\Psi_2$ is then ruled
approximately by:
\begin{eqnarray}
  \label{eq:TRsimple}
  \lefteqn{{\overline\rho}\frac{L}{M}\left\{\frac{2}{3}\left[1-\frac{{\overline\Omega}^2}{2\pi G{\overline \rho}}\right]{\overline\Omega}^2\partial_{r}\left(\frac{r^2}{\overline g}\right)-\frac{2}{9}\frac{\rho_m}{\overline\rho}\frac{r^2}{\overline g}\partial_r{\overline\Omega}^2\right.}\nonumber\\
  &+&{\left.\frac{\rho_m}{\overline\rho}\left[\frac{r}{3}\partial_{r}\left[H_T\partial_r\Psi_2\!-\!\left(1-\delta+\chi_T\right)\Psi_2\right]
        \!-\!\frac{2 H_T}{3 r}\left(1+\frac{D_h}{K_T}\right)\Psi_2\right]\right\}}\nonumber\\
  &\approx&{\overline \rho}{\overline
    T}C_{p}\frac{N_{T}^{2}}{{\overline g}\delta}U_{2}.
\end{eqnarray}
Note that $\tilde{\mathcal T}_{{\mathcal B}4}$ is now expressed directly as
a function of ${\overline\Omega}$ where Eq. (\ref{ThermalWind}) has been
used. On the other hand, we have verified that the $\Lambda_2$-terms can be
neglected compared to the $\Psi_2$ ones during the main sequence evolution
which allows simplifications in the writing of $\tilde{\mathcal T}_{{\rm
    Th}1}$. The advection of entropy is then balanced by $\tilde{\mathcal
  T}_{{\mathcal B}1}$, $\tilde{\mathcal T}_{{\mathcal B}4}$ and the thermal
diffusion. The simplification given by Eq.~(\ref{eq:TRsimple}) is no longer
valid if structural readjustments occur, as it is the case when the core
contracts at the end of the main sequence.

If \citet{zahn92} prescription for $D_h$ is used, the term
$\frac{D_h}{K_T}$ can be furthermore omitted.

\vspace{1em}

n conclusion, the picture of the heat transport presented
here differs radically from the classical Eddington-Sweet vision
where the thermal imbalance induced by the centrifugal acceleration
is considered to be at the origin of the meridional circulation.
Starting from an arbitrary rotation and thermal state, the advection of
entropy leads, after thermal relaxation, to a new temperature
state and thus to a new differential rotation profile (cf. Eq.\ref{ThermalWind}).
There then remains a circulation induced by the angular momentum
losses and the structural adjustments, and the concomitant evolution of
the rotation and thermal states.

\subsection{Transport of chemicals}

Once a composition gradient  builds up, due to gravitational settling (not 
accounted for in the present calculations) or
nuclear burning, the meridional circulation creates a latitudinal
perturbation of the molecular weight distribution, as described by
Eq.~(\ref{Lambda}). The sign of these $\mu$-fluctuations depends on the
orientation of the circulation, and we thus expect some differences between
the two models.

\begin{figure}[tbp]
\begin{center}
\includegraphics[width=0.35\textwidth]{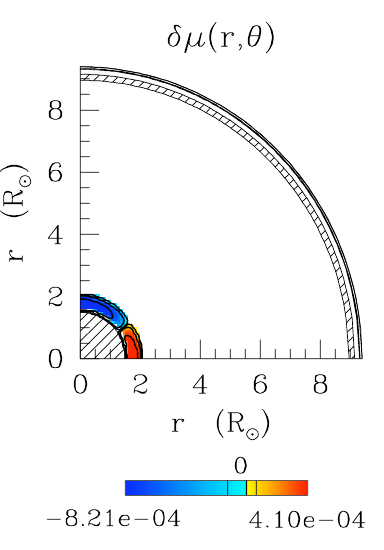}
\caption{Two-dimensional reconstruction of the $\mu$-perturbations,
  $\delta\mu = \overline\mu\Lambda_2 P_2\left(\cos\theta\right)$, in the
  20~\Ms{} star when $X_\text{c} = 0.32$. Hatched regions delineate
  convection zones.}
\label{fig:2Dmu20}  
\end{center}
\end{figure}

\begin{figure}[tbp]
\begin{center}
\includegraphics[width=0.35\textwidth]{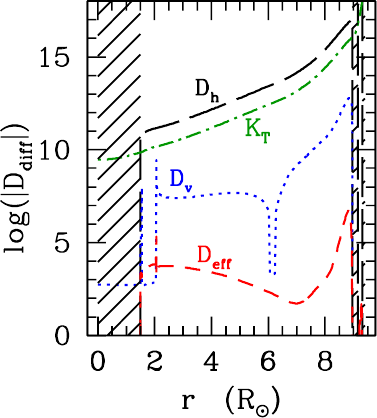}
\caption{Profiles of the thermal diffusivity ($K_T$ dotted-dashed cyan
  line) and that associated to the meridional circulation ($D_{\rm eff}$
  short-dashed red line) and with the horizontal ($D_h$ long-dashed black
  line) and vertical ($D_v$ dotted blue line) turbulence in the 20~\Ms{}
  model when $X_\text{c}=0.32$.}
\label{fig:diff20}
\end{center}
\end{figure}

\subsubsection{20 M$_{\odot}$ model}

In that model, the meridional circulation always transports matter with
low-$\mu$ down to the interior near the rotational axis, while high-$\mu$
matter is carried outward along the equator. As a result, a horizontal
$\mu$-gradient appears near the convective core (see
Fig.~\ref{fig:2Dmu20}), which is directed from pole to equator,
i.e. opposite to that of the temperature fluctuations. These
$\mu$-fluctuations remain small (less than 1\%), which justifies the
perturbative approach.

We already mentioned in \S~\ref{turb-mod} that the strong turbulent
transport in the horizontal direction interferes with the meridional
circulation, rendering the vertical transport of diffusive nature,
characterized by an effective diffusivity $D_\text{eff}$ \citep{CZ92}. 
This description makes the comparison between the two processes that contribute to the
transport of chemical elements (vertical shear and combination of advection
and horizontal shear) easier since it suffices to compare their respective
diffusivities.

This is done in Fig.~\ref{fig:diff20}, where we see that in most of the
stellar interior the vertical shear turbulence largely dominates over the
combination of advection and horizontal shear ($D_v \gg D_\text{eff}$),
except in the vicinity of the convective core, where the strong
$\mu$-stratification inhibits the shear instability, and in the region
where the gradient of $\overline \Omega$ changes sign (and where $D_v$
reduces to the molecular diffusivity).

This figure can be compared with Fig. 6 of \citet{MM00}, where the same
profiles are presented for a 20~\Ms{} star at Z = 0.02, with initial
equatorial velocity $\upsilon_{\rm ini} = 300$~\kms. In their simulation,
Meynet \& Maeder use the prescription given by Eq.~(\ref{nuhz}) for the
horizontal turbulent viscosity. While we obtain very similar profiles for
the vertical shear diffusion coefficient, $D_v$, and for the thermal
diffusivity, $K_T$, the diffusion coefficients associated with horizontal
turbulence, $D_h$, and meridional circulation, $D_{\rm eff}$, differ by
several orders of magnitude. The \citet{MPZ04} prescription results in
larger horizontal turbulent viscosity, which also translates into lower
$D_{\rm eff}$. This very same result was firstly obtained by
\citet{Maeder03}, where using energetic considerations, he derived a new
expression for $D_h$ that is very similar to Eq.~\ref{nuhg} \citep[see
also][]{MPZ07}. It should be noted that the anisotropic turbulence
assumption ($D_h \gg D_v$) is always satisfied.

\subsubsection{1.5 M$_{\odot}$ model}

\begin{figure*}[tbp]
\begin{center}
\includegraphics[width=0.925\textwidth]{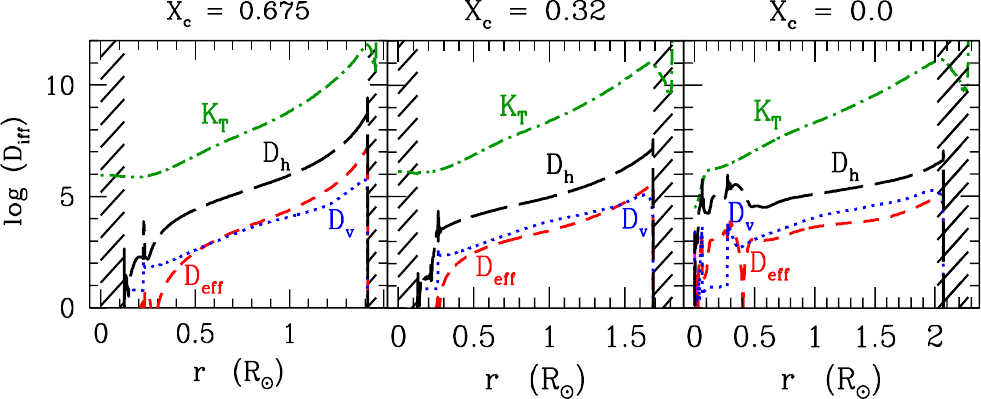}
\caption{Same as Fig.~\ref{fig:diff20} for the 1.5~\Ms{} model when
  $X_\text{c} = 0.675$, $X_\text{c} = 0.32$ and $X_\text{c} = 0.0$.}
\label{fig:diff1.5}
\end{center}
\end{figure*}

\begin{figure*}[tbp]
\centering
\includegraphics[width=0.925\textwidth]{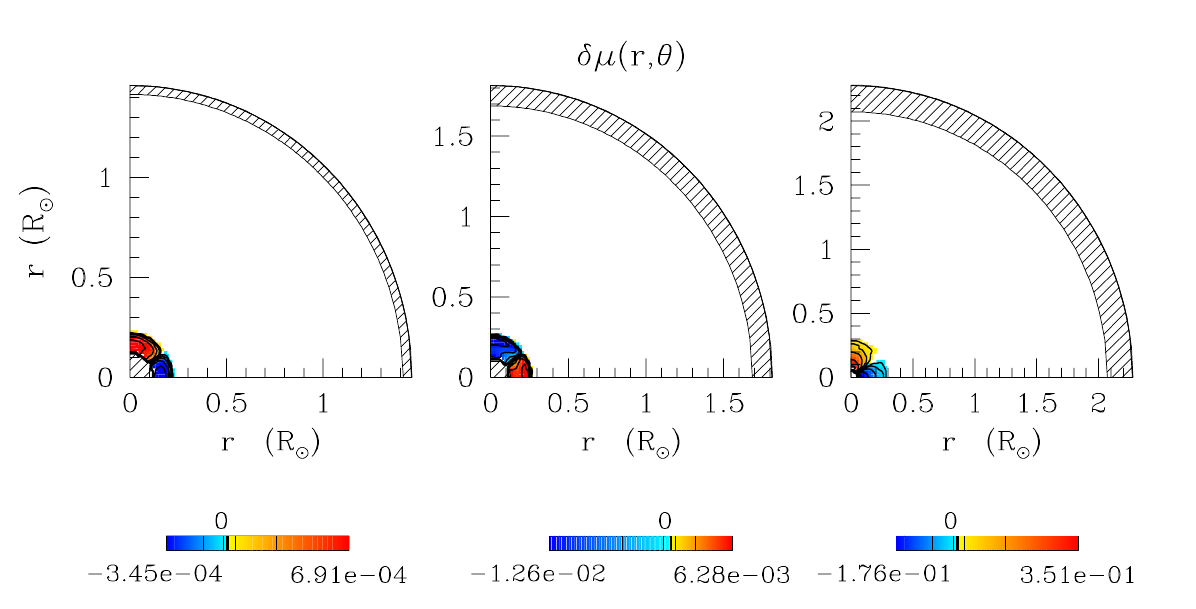}
\caption{Two-dimensional reconstruction of the $\mu$-perturbations,
  $\delta\mu=\overline\mu\Lambda_2 P_2\left(\cos\theta\right)$ in the
  1.5~\Ms{} when $X_\text{c} = 0.675$ (\emph{left panel}), $X_\text{c} =
  0.32$ (\emph{middle panel}) and $X_\text{c} = 0.0$ (\emph{right
    panel}). Hatched regions delineate convection zones.}
\label{fig:2Dmu1.5}
\end{figure*}

Figure~\ref{fig:diff1.5} shows that the diffusion coefficients in the 1.5
M$_{\odot}$ star  are much smaller than in the massive star as a result of
lower angular velocity and degree of differential rotation. The
  diffusion coefficients $D_\text{eff}$ and $D_v$ are much closer in this
  model, partly due to the adoption of \citeauthor{zahn92}'s \citeyearpar{zahn92} prescription for $D_h$
  (see Eqs.~(\ref{nuhg}) and (\ref{nuhz}) and Mathis et al. 2004 for more
  details). However, the decreasing efficiency of the meridional transport
  as the star evolves on the main sequence
  results in the shear turbulence dominating the distribution of
  chemicals \citep[see][]{PA03}. The inhibiting action of the composition gradient is visible
  near the convective core where $D_v$ drops to its microscopic value,
  indicating that turbulence is suppressed.  It should also be noted that
  in this model  the anisotropic turbulence assumption ($D_h \gg D_v$) is
  also always satisfied.

The sign of the $\mu$-perturbation depends on the orientation of the
circulation near the convective core. In the lower mass model,
differences arise from the development of a secondary loop, as seen at
$X_\text{c}=0.675$ and $X_\text{c}=0$.

Although the meridional circulation and $D_{h}$ are both smaller in the
low-mass star, the ratio $U_2/D_{h}$, which governs the relative strength
of vertical advection by meridional circulation and of horizontal diffusion
(see Eq. \ref{Lambda}), is larger in the 1.5\Ms{} compared to the
20~\Ms{}. This results in larger $\mu$-imbalance in the lower-mass model.

\section{Conclusion}

The paper presents a complete set of diagnostic tools that
provide a comprehensive and coherent understanding of the secular
hydrodynamical transport processes operating in the radiative zones of
rotating stars. The framework is that of ``type I rotational mixing'', where
angular momentum and nuclides are transported by the same mechanisms,
namely large-scale meridional circulation and shear-induced turbulence.

\begin{figure}[tbp]
\begin{center}
\includegraphics[width=0.45\textwidth]{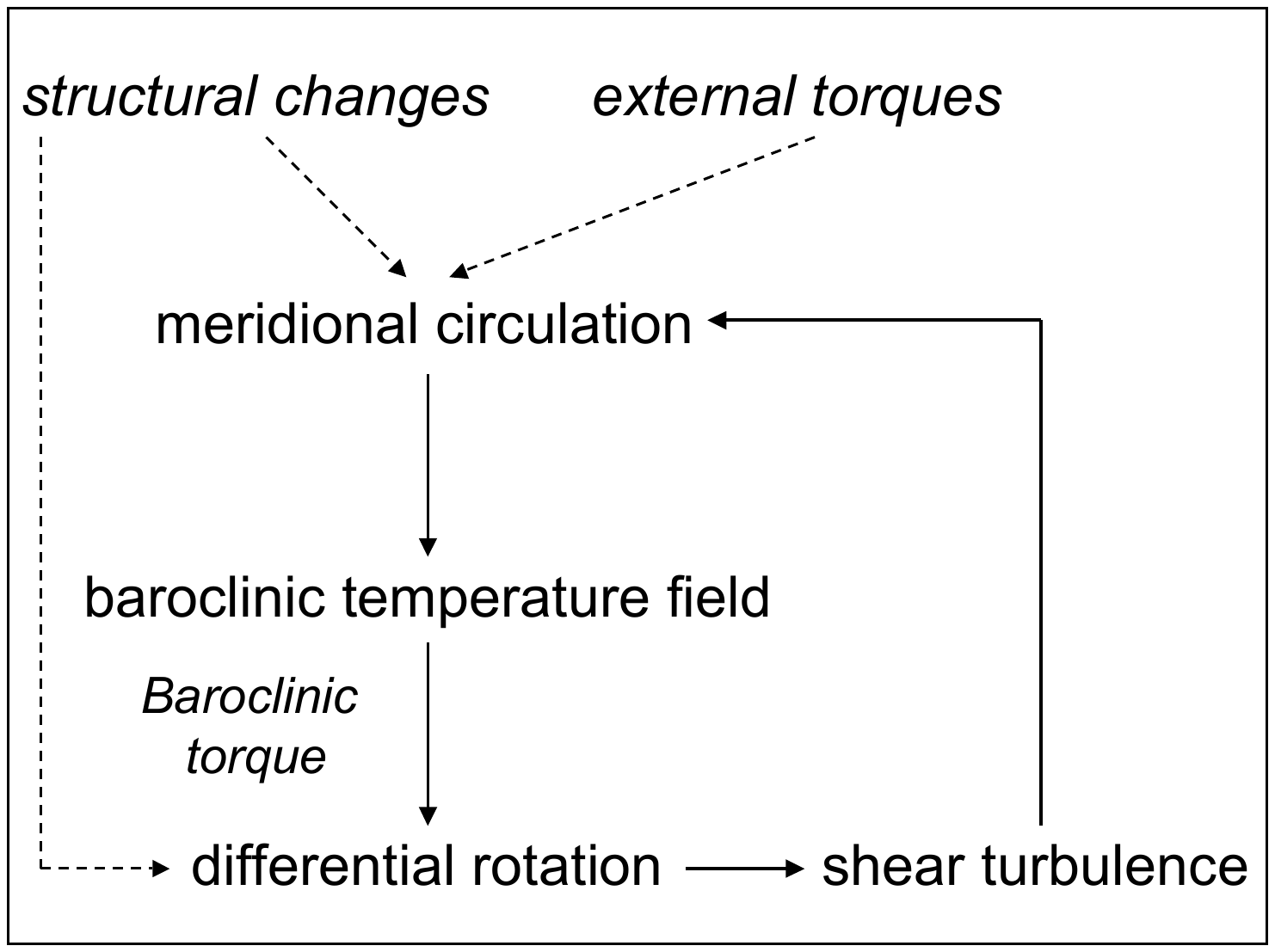}
\caption{Rotational mixing of type I in the radiative zone of a rotating
  star, where the transport of angular momentum is achieved by meridional
  circulation and turbulent diffusion.}
\label{fig:loopth}
\end{center}
\end{figure}

In order to validate our new approach, the first analysis presented here 
is performed on two well-studied cases of a low-mass
(1.5~\Ms{}) \citep[see][]{PA03} and a massive (20~\Ms{}) star
\citep{MM00}. It allows us to disentangle what are the main processes at the origin of 
meridional circulation and driving angular momentum, heat and chemicals transports 
in the stellar interior
and to confirm results previously established in the literature.
In particular, the angular momentum loss, either by radiation-driven wind (the massive
star) or by magnetic braking (the low-mass star) combined with
structural readjustments are the sources responsible for the generation of
meridional circulation (Fig.~\ref{fig:loopth}). The direction of the flow
is such that it transports angular momentum outward to the surface, except
in the deep interior of the 1.5~\Ms{} star where it can reverse, producing
a steeper angular velocity gradient.

This circulation advects heat, and thus generates latitude dependent
temperature fluctuations; these tend to be damped out by radiative
diffusion, until they establish a subtle balance between advection and
diffusion, that allows them to induce differential rotation through the
baroclinic torque (Eq. \ref{ThermalWind}). This differential rotation is
shear-unstable, and generates turbulence that participates to the transport
of angular momentum, thus closing the loop (Fig.~\ref{fig:loopth}). This
picture thus differs drastically from the classical Eddington-Sweet circulation, where
the circulation is deemed to originate from the thermal imbalance due to
the centrifugal force. If turbulent transport and extraction of angular
  momentum were absent, the circulation would die altogether, as
was pointed out by \citet{busse82}.

Thanks to the diagnostic tools we have developed, the present investigation
helped to clarify a number of other points. We confirm that it is mainly
the meridional circulation that transports angular momentum in the low-mass
star \citep[see also][]{PA03}; it is thus crucial to describe that
transport as an advective process, as in the approach developed by Zahn, Maeder, 
and collaborators. In the massive star, where the
redistribution of angular momentum due to the star's expansion plays a much
more important role than its extraction by the radiative wind, advection
and diffusion make comparable contributions, though in different regions:
the advective transport dominates in the inner layers of the star, whereas
the diffusive transport is the prime actor in the upper part and, moreover,
works there in the opposite direction.

For the first time, we
confirm through detailed computation and separation of the different
components, that the advection of entropy by meridional circulation is
almost exactly balanced by the thermal relaxation (barotropic and thermic
terms) during most of the main sequence. This was generally assumed prior
to the \citet{MZ98} work integrating the time-dependency of the
temperature fluctuations, but not actually verified through the
computation. Grounding our detailed analysis on advanced graphical
tools, we are also able to propose a simplified expression for the thermal
relaxation on the main sequence.

On the other hand, the transport of
chemicals is shown to be dominated by the shear-induced turbulence while
the star evolves on the main sequence. This is particularly clear for a
massive star such as the 20~\Ms{} star.

In the near future, we plan to apply the same tools to Type II rotational
mixing, where we shall include the transport of angular momentum by
internal gravity waves and magnetic stresses and to advanced phases of
stellar evolution.

\begin{acknowledgements}
  We thank the referee A. Maeder for useful remarks that helped improving
  the final version of this paper.  This work received financial support
  from the Programme National de Physique Stellaire of CNRS/INSU (France).
  T.D. and C.C. acknowledge financial support from Swiss FNS. LS is FNRS
  Research Associate.  This research has made use of NASA's Astrophysics
  Data System Bibliographic Services.
\end{acknowledgements}


\appendix

\section{Meridional circulation and angular momentum transport}\label{ap:A1}

Let us consider the Eq. (\ref{AM}) for the angular momentum transport: 
\begin{equation}
  \overline\rho\frac{{\rm d}}{{\rm d}t}\left(r^2\overline{\Omega}\right)=\frac{1}{5r^2}\partial_{r}\left(\overline\rho r^4 \overline{\Omega}U_{2}\right)+\frac{1}{r^2}\partial_{r}\left(\overline\rho\nu_{v}r^4\partial_{r}\overline{\Omega}\right)\, .
\end{equation}
Integrating over a spherical shell, we get:
\begin{eqnarray}
\lefteqn{\int_{r=r_1}^{r=r_2}\overline\rho {r'}^{2}\frac{\rm d}{{\rm d}t}\left({r'}^{2}\overline{\Omega}\right){\rm d}{r'}=\frac{1}{5}\int_{r=r_1}^{r=r_2}\partial_{r'}\left(\overline\rho {r'}^{4} \overline{\Omega} U_{2}\right){\rm d}{r'}}\nonumber\\
&+&\int_{r=r_1}^{r=r_2}\partial_{r'}\left(\overline\rho\nu_{v}{r'}^4\partial_{r}\overline{\Omega}\right){\rm d}{r'}\, ,
\end{eqnarray}
$r_{1}$ and $r_{2}$ being respectively its inner and outer radius. Then, introducing the elementary mass element ${\rm d}m=4\pi\overline\rho r^2 {\rm d}r$, we obtain:
\begin{eqnarray}
\lefteqn{\frac{1}{4\pi}\int_{m=m\left(r_1\right)}^{m=m\left(r_2\right)}\frac{\rm d}{{\rm d}t}\left({r'}^{2}\overline{\Omega}\right){\rm d}m=\frac{1}{5}\left[\left(\overline\rho r^4 \overline{\Omega} U_{2}\right)_{r=r_2}-\left(\overline\rho r^4 \overline{\Omega} U_{2}\right)_{r=r_1}\right]}\nonumber\\
&+&\left[\left(\overline\rho\nu_{v}r^4\partial_{r}\overline{\Omega}\right)_{r=r_2}-\left(\overline\rho\nu_{v}r^4\partial_{r}\overline{\Omega}\right)_{r=r_1}\right]\, .
\end{eqnarray}
We apply the following identity
\begin{eqnarray}
\lefteqn{\frac{\rm d}{{\rm d}t}\left[\int_{m_1\left(t\right)}^{m_2\left(t\right)}f(m,t){\rm d}m\right]=\int_{m_1\left(t\right)}^{m_2\left(t\right)}\frac{\rm d}{{\rm d}t}\left[f(m,t)\right]{\rm d}m}\nonumber\\
&+&\left\{\frac{{\rm d}m_2\left(t\right)}{{\rm d}t}\cdot f\left(m_2,t\right)-\frac{{\rm d}m_1\left(t\right)}{{\rm d}t}\cdot f\left(m_1,t\right)\right\}.
\end{eqnarray}
In our Lagrangian description, the mass is conserved, $\frac{{\rm
    d}m_2\left(t\right)}{{\rm d}t}=\frac{{\rm d}m_1\left(t\right)}{{\rm
    d}t}=0$, so
\begin{eqnarray}
\lefteqn{\frac{1}{4\pi}\frac{{\rm d}}{{\rm d}t}\left[\int_{m=m\left(r_1\right)}^{m=m\left(r_2\right)}r'^{2}\overline{\Omega}{\rm d}m\right]=\frac{1}{5}\left[\left(\overline\rho r^4 \overline{\Omega} U_{2}\right)_{r=r_2}-\left(\overline\rho r^4 \overline{\Omega} U_{2}\right)_{r=r_1}\right]}\nonumber\\
&+&\left[\left(\overline\rho\nu_{v}r^4\partial_{r}\overline{\Omega}\right)_{r=r_2}-\left(\overline\rho\nu_{v}r^4\partial_{r}\overline{\Omega}\right)_{r=r_1}\right]\, .
\label{Bilan}
\end{eqnarray}
To obtain the value of the meridional circulation at $m(r)$, we set $r_{1}=0$ and
$r_{2}=r$\,:
\begin{equation}
U_{2}=\frac{5}{\overline \rho r^4 \overline{\Omega}}\left[\Gamma\left(m\right)-\overline\rho\nu_{v}r^4\partial_{r}\overline{\Omega}\right]
\end{equation}
where
\begin{equation}
\Gamma\left(m\right)=\frac{1}{4\pi}\frac{\rm d}{{\rm d}t}\left[\int_{0}^{m\left(r\right)}{r'}^2\overline{\Omega}{\rm d}m'\right].
\end{equation}

\section{Components ${\mathcal T}_2$ of the heat equation}\label{ap:A2}

These terms intervene in the heat relaxation equation (\ref{eq:Mer}) and
their derivation may be found in Zahn (1992) or Mathis \& Zahn (2004). They
include : 

\noindent the barotropic term:
\begin{eqnarray}
\lefteqn{{\mathcal T}_{2,{\mathcal B}}=}\nonumber\\
&&\frac{2}{3}\left[1-\frac{\overline\Omega^2}{2\pi G\overline{\rho}}-\frac{2}{3}\frac{\rho_m}{\rho}\left(\varphi\Lambda_{2}-\delta\Psi_{2}\right)-\frac{\left(\overline{\epsilon}+\overline{\epsilon}_{\rm grav}\right)}{\epsilon_{m}}\right]\overline\Omega^{2}\partial_{r}\left(\frac{r^2}{\overline g}\right)\nonumber\\
&&-\frac{2}{3}\frac{\rho_m}{\overline\rho}\left(\varphi\Lambda_{2}-\delta\Psi_{2}\right),
\label{TB}
\end{eqnarray}

\noindent the thermic term:
\begin{eqnarray}
{\mathcal T}_{2,{\rm Th}}=\frac{\rho_{m}}{\overline{\rho}}\left[\frac{r}{3}\partial_{r}{\cal{A}}_{2}\left(r\right)
-\frac{2 H_{T}}{3 r}\left(1 +
\frac{D_{h}}{K_{T}}\right){\Psi_{2}}\right],
\label{Tth}
\end{eqnarray}

\noindent the term associated with local energy sources:
\begin{eqnarray}
{\mathcal T}_{2,{\rm N-G}}&=&\frac{\left(\overline{\epsilon}+\overline{\epsilon}_{\rm grav}\right)}{\epsilon_{m}}\left[{\cal{A}}_{2}\left(r\right)+(f_{\epsilon}\epsilon_{T} - f_{\epsilon}\delta + \delta)\Psi_{2}\right.\nonumber\\
&&{\left.+(f_{\epsilon}\epsilon_{\mu}+f_{\epsilon}\varphi - \varphi)\Lambda_{2}\right]}\nonumber
\end{eqnarray}
where
\begin{eqnarray}
{\cal{A}}_{2}\left(r\right)=H_{T}\partial_{r}\Psi_{2}-(1-\delta+\chi_{T})\Psi_{2}-(\varphi+\chi_{\mu})\Lambda_{2}.
\end{eqnarray}
$N_{T}^{2}=({{\overline g}\delta}/{H_{P}})\left(\nabla_{\rm
    ad}-\nabla\right)$ is the square of the thermal part of the Br\"{u}nt-Va\"{\i}s\"{a}l\"{a}
frequency where $\nabla=\partial\ln T/\partial\ln P$ is the radiative
temperature gradient and $\nabla_{\rm ad}$ is the adiabatic one. $L$, $M$,
$C_P$ have their usual meaning and $\overline{T}$ is the horizontal average
of the temperature. We have also introduced the pressure and 
temperature scale-heights $H_{P}=\left|{\rm d}r/{\rm d}\ln P\right|$ and
$H_{T}=\left|{{\rm d}r}/{{\rm d}\ln \overline{T}}\right|$, the thermal
diffusivity
$K_{T}={\overline{\chi}}/{\overline{\rho}C_{P}}= 16 \sigma T^3/3 \kappa {\overline{\rho}^2C_{P}}$
 the
horizontal eddy-diffusivity $D_{h}$ and
$f_{\epsilon}={\overline{\epsilon}}/{\left(\overline{\epsilon}+\overline{\epsilon}_{\rm
      grav}\right)}$, with $\overline{\epsilon}$ and
$\overline{\epsilon}_{\rm grav}$ being the mean nuclear and gravitational energy
release rates respectively. $\sigma$ and $\kappa$ are  the
Stefan constant and the Rosseland mean
opacity. $\epsilon_{\mu}=\left(\partial\ln\epsilon/\partial\ln\mu\right)_{P,T}$
and $\chi_{\mu}=\left(\partial\ln\chi/\partial\ln \mu\right)_{P,T}$ are the
logarithmic derivatives of $\epsilon$ and of the radiative conductivity
$\chi$ with respect to $\mu$, and
$\epsilon_{T}=\left(\partial\ln\epsilon/\partial\ln T\right)_{P,\mu}$ and
$\chi_{T}=\left(\partial\ln\chi/\partial\ln T\right)_{P,\mu}$ are the
derivatives of these same quantities with respect to the temperature
$T$. Finally $\epsilon_{m}=L/M$ and $\rho_{m}={3{\overline
    g}\left(r\right)}/{4\pi r G}$ are the horizontal average of the energy production rate and
the mean density
inside the considered level surface.


\begin{thebibliography}{}


\bibitem[Abt et al.(2002)]{Abt02} Abt, H.~A., Levato, H., 
\& Grosso, M. 2002, ApJ, 573, 359 

\bibitem[Bouret et al.(2008)]{Bouret08} Bouret, J.~-C., Donati, 
J.~-F., Martins, F., Escolano, C., Marcolino, W., Lanz, T., 
\& Howarth, I. 2008, ArXiv e-prints, 806, arXiv:0806.2162 

\bibitem[Braithwaite(2006)]{B06}
Braithwaite, J. 2006, A\&A, 449, 451

\bibitem[Braithwaite \& Spruit(2005)]{BS05}
Braithwaite, J., \& Spruit, H. 2005, Nature, 431, 819

\bibitem[Brun \& Zahn(2006)]{BZ06}
Brun, A.-S., \& Zahn, J.-P. 2006, A\&A, 457, 665

\bibitem[Busse(1982)]{busse82}
Busse, F. H. 1982, ApJ, 259, 759

\bibitem[Chaboyer \& Zahn(1992)]{CZ92}
Chaboyer, B., \& Zahn, J.-P. 1992, A\&A, 253, 173

\bibitem[Charbonnel \& Talon(2005)]{CT05}
Charbonnel, C., \& Talon, S. 2005, Science, 309, 2189

\bibitem[Charbonneau \& MacGregor(1993)]{CMG93}
Charbonneau, P.,\&  MacGregor, K. B. 1993, ApJ, 417, 762

\bibitem[Decressin et al.(2007)]{DMCPE07} 
Decressin, T., Meynet, G., Charbonnel, C., Prantzos, N., \& Ekstr{\"o}m, S.  2007, A\&A, 464, 1029

\bibitem[Eddington(1925)]{E25}
Eddington, A. S., 1925, Obs, 48, 73

\bibitem[Ekstr\"om et al.(2008)]{E08}
Ekstr\"om, S., Meynet, G., Maeder, A., Barblan, F., 2008, A\&A, 478, 467

\bibitem[Eggenberger et al.(2005)]{Egg05}
Eggenberger, P., Maeder, A., \& Meynet, G. 2005, A\&A, 440, L9

\bibitem[Endal \& Sofia(1978)]{Endal78} Endal, A.S., Sofia, S. 1978, ApJ, 220, 279

\bibitem[Espinosa Lara \& Rieutord(2007)]{ER07} Espinosa Lara, F., \& Rieutord, M. 2007, A\&A, 470, 1013

\bibitem[Fukada(1982)]{F82} Fukada, I. 1982, PASP, 94, 271

\bibitem[Gaig\'e(1993)]{G93} Gaig\'e, Y. 1993, A\&A, 269, 267 

\bibitem[Garaud(2002a)]{Garaud02a}
Garaud, P. 2002a, MNRAS, 329, 1

\bibitem[Garaud(2002b)]{Garaud02b}
Garaud, P. 2002b, MNRAS, 335, 707

\bibitem[Grevesse et al.(1996)]{GNS96} 
Grevesse, N., Noels, A., \& Sauval, A.~J. 1996, Cosmic Abundances, 99, 117

\bibitem[Heger et al.(2000)]{Heger00} Heger, A., Langer, N., \& Woosley, S. E. 2000, ApJ, 528, 368

\bibitem[Heyney et al.(1964)]{Heyney64} Henyey, L. G., Forbes, J. E., \& Gould, N. L. 1964, ApJ, 139, 306  

\bibitem[Kawaler(1988)]{K88}
Kawaler, S. D. 1988, ApJ, 333, 236

\bibitem[Kippenhahn \& Weigert(1990)]{KW90}
Kippenhahn, R., \& Weigert, A. 1990, {\it Stellar Structure and Evolution}, Springer-Verlag

\bibitem[Maeder(1999)]{M99}
Maeder, A. 1999, A\&A, 347, 185

\bibitem[Maeder(2002)]{Maeder02}
Maeder, A. 2002, A\&A, 392, 575

\bibitem[Maeder(2003)]{Maeder03}
Maeder, A. 2003, A\&A, 399, 263

\bibitem[Maeder(2009)]{M09}
Maeder, A., 2009, {\it Physics, Formation and Evolution of
Rotating Stars}, Springer-Verlag

\bibitem[Maeder \& Zahn(1998)]{MZ98}
Maeder, A., \& Zahn, J.-P. 1998, A\&A, 334, 1000


\bibitem[Maeder \& Meynet(2000a)]{MaMe00}
Maeder, A., \& Meynet, G.\ 2000a, \aap, 361, 159

\bibitem[Maeder \& Meynet(2000b)]{MM00b}
Maeder, A., \& Meynet, G.\ 2000b, ARAA, 38, 143

\bibitem[Maeder \& Meynet(2001)]{MM01} 
Maeder, A., \& Meynet, G.\ 2001, \aap, 373, 555

\bibitem[Maeder \& Meynet(2004)]{MM04}
Maeder, A., \& Meynet, G. 2004, A\&A, 422, 225

\bibitem[Mathis \& Zahn(2004)]{MZ04}
Mathis, S., \& Zahn, J.-P. 2004, A\&A, 425, 229

\bibitem[Mathis \& Zahn(2005)]{MZ05}
Mathis, S., \& Zahn, J.-P. 2005, A\&A, 440, 653

\bibitem[Mathis et al.(2004)]{MPZ04}
Mathis, S., Palacios, A., \& Zahn, J.-P. 2004, A\&A, 425, 243

\bibitem[Mathis et al.(2007)]{MPZ07}
Mathis, S., Palacios, A., \& Zahn, J.-P. 2007, A\&A, 462, 1063

\bibitem[Mathis et al.(2008)]{Mathis08} Mathis, S., Talon, S., Pantillon, F.-P., \& Zahn, J.-P. 2008, \solphys, in press (doi: 10.1007/s11207-008-9157-0)

\bibitem[Menou et al.(2004)]{Menou04}
Menou, K., Balbus, S. A., \& Spruit, H. C. 2004, ApJ, 607, 564

\bibitem[Mestel(1953)]{M53}	
Mestel, L., 1953, MNRAS, 113, 716

\bibitem[Meynet \& Maeder(2000)]{MM00}
Meynet, G., \& Maeder, A. 2000, A\&A, 361, 101

\bibitem[Meynet et al.(2007)]{M07}
Meynet, G., Ekstr\"om, S., Maeder, A., Barblan, F., 2007, ASP, 361, 325

\bibitem[Palacios et al.(2003)]{PA03}
Palacios, A., Talon, S., Charbonnel, C., \& Forestini, M. 2003, A\&A, 399, 603

\bibitem[Palacios et al.(2006)]{PCTS06}
Palacios, A., Charbonnel, C., Talon, S., \& Siess, L. 2006, A\&A, 453, 261

\bibitem[Pantillon et al.(2007)]{Pantillon07} Pantillon, F.P., Talon, S., \& Charbonnel, C. 2007, A\&A, 474, 155

\bibitem[Pinsonneault et al.(1989)]{Pinsonneault89} 
Pinsonneault, M.H., Kawaler, S.D., Sofia, S., \& Demarque,
  P. 1989, ApJ, 338, 424

\bibitem[Press(1981)]{Press81}
Press, W.-H. 1981, ApJ, 245, 286


\bibitem[Reimers(1975)]{Reimers75} 
Reimers, D. 1975, M\'emoires de la Soci\'et\'e Royale des Sciences de Li\`ege, 8, 369 

\bibitem[Rieutord(2006a)]{Rieutord06}
Rieutord, M. 2006a, {\it Stellar Fluid Dynamics and Numerical Simulations: From the Sun to Neutron Stars}, eds. M. Rieutord and B. Dubrulle, EAS, 21, 275

\bibitem[Rieutord(2006b)]{Rieutord06b} Rieutord, M. 2006b, A\&A, 451, 1025

\bibitem[Rogers \& Glatzmaier(2005)]{RG05}
Rogers, T., Glatzmaier, G. 2005, MNRAS, 364, 1135

\bibitem[Schatzman(1962)]{Schatzman62}
Schatzman, E., 1962, Annales d'Astrophysique, 25, 18

\bibitem[Schatzman(1993)]{Schatzman93}
Schatzman, E. 1993, A\&A, 279, 431 

\bibitem[Schnerr et al.(2008)]{Schnerr08} 
Schnerr, R.~S., et al. 2008, A\&A, 483, 857 

\bibitem[Spruit(1999)]{Spruit99}
Spruit, H. C. 1999, A\&A, 349, 189

\bibitem[Spruit(2002)]{Spruit02}
Spruit, H. C. 2002, A\&A, 381, 923

\bibitem[Siess(2006)]{Siess06}
Siess, L. 2006, A\&A, 448, 717

\bibitem[Siess et al.(2000)]{Siess00}
Siess, L., Dufour, E., \& Forestini, M. 2000, A\&A, 358, 593

\bibitem[Sweet(1950)]{S50}
Sweet, P. A., 1950, MNRAS, 110, 548

\bibitem[Talon(1997)]{Talon97}
Talon, S. 1997, Ph.D. Thesis

\bibitem[Talon(2007)]{talon07}
Talon, S. 2007, Transport processes in stars: diffusion, rotation, magnetic
fields and internal waves in Stellar Nucleosynthesis: 50 years after BBFH, eds. C. Charbonnel and J.-P. Zahn, EAS publication series, to appear ({arXiv:0708.1499})

\bibitem[Talon \& Zahn(1997)]{TZ97}
Talon, S., \& Zahn, J.-P. 1997, A\&A, 317, 749

\bibitem[Talon \& Charbonnel(1998)]{TC98} 
Talon, S., \& Charbonnel, C. 1998, A\&A, 335, 959 

\bibitem[Talon \& Charbonnel(2003)]{TC03}
Talon, S., \& Charbonnel, C. 2003, A\&A, 405, 1025

\bibitem[Talon \& Charbonnel(2004)]{TC04}
Talon, S., \& Charbonnel, C. 2004, A\&A, 418, 1051

\bibitem[Talon \& Charbonnel(2005)]{TC05}
Talon, S., \& Charbonnel, C. 2005, A\&A, 440, 981

\bibitem[Talon \& Charbonnel(2008)]{TC08}
Talon, S., \& Charbonnel, C. 2008, Proc. IAU Symp. 252 ({arXiv:0805.4697})


\bibitem[Talon et al.(2002)]{TKZ02}
Talon, S., Kumar, P., \& Zahn, J.-P. 2002, ApJ, 574, L175

\bibitem[Talon et al.(1997)]{TZMM97} 
Talon, S., Zahn, J.-P., Maeder, A., \& Meynet, G.\ 1997, A\&A, 322, 209

\bibitem[Vink et al.(2000)]{Vink00} 
Vink, J.~S., de Koter, A., \& Lamers, H.~J.~G.~L.~M. 2000, A\&A, 362, 295 

\bibitem[Vogt(1925)]{V25}
Vogt, H., 1925,	AN, 223, 229

\bibitem[Zahn(1992)]{zahn92}
Zahn, J.-P. 1992, A\&A, 265, 115

\bibitem[Zahn et al.(2007)]{ZBM07} Zahn, J.-P., Brun, A.~S., \& Mathis, S.\
  2007, A\&A, 474, 145 


\end{thebibliography}
\end{document}